\documentclass[sigconf]{acmart}
\AtBeginDocument{%
  \providecommand\BibTeX{{%
    \normalfont B\kern-0.5em{\scshape i\kern-0.25em b}\kern-0.8em\TeX}}}

\setcopyright{acmcopyright}
\copyrightyear{2018}
\acmYear{2018}
\acmDOI{10.1145/1122445.1122456}

\usepackage{booktabs} % For formal tables
\usepackage{url}

\usepackage{caption}
\usepackage[all]{nowidow}
\usepackage{wrapfig}
\usepackage{array}
\usepackage{arydshln}
\usepackage{tabularx}
\usepackage{multirow}
\usepackage{arydshln}
\newcolumntype{L}[1]{>{\raggedright\let\newline\\\arraybackslash\hspace{0pt}}m{#1}}
\newcolumntype{C}[1]{>{\centering\let\newline\\\arraybackslash\hspace{0pt}}m{#1}}
\newcolumntype{R}[1]{>{\raggedleft\let\newline\\\arraybackslash\hspace{0pt}}m{#1}}

\usepackage{wrapfig,lipsum,booktabs} % for wrapping table

\def\authnotes{1}
\newcounter{notectr}[section]
\newcommand{\thenote}{\thesubsection.\arabic{notectr}\refstepcounter{notectr}}
\newcommand{\newedits}[1]{\textcolor{black}{#1}}
\newcommand{\note}[2]{$\ll$#1~\thenote: #2$\gg$}
\newcommand{\cnote}[1]{\ifnum\authnotes=1 \textcolor{blue}{\note{Comment:}{#1}}\fi}

\copyrightyear{2025}
\acmYear{2025}
\setcopyright{acmlicensed}\acmConference[DIS'25]{DIS}{Jun 5-9, 2025}{Japan}
\acmBooktitle{DIS'24, Apr 5-9, 2025, Portugal}
\acmPrice{15.00}
\acmDOI{10.1145/3491102.XXXXXX}
\acmISBN{978-1-XXXXXXX}

\copyrightyear{2025}
\acmYear{2025}
\acmConference[DIS '25]{Designing Interactive Systems Conference}{July 5--9,2025}{Funchal, Portugal}
\acmBooktitle{Designing Interactive Systems Conference (DIS '25), July 5--9,
2025, Funchal, Portugal}\acmDOI{10.1145/3715336.3735725}
\acmISBN {979-8-4007-1485-6/2025/07}

\begin{document}

%%
%% The "title" command has an optional parameter,
%% allowing the author to define a "short title" to be used in page headers.

%\title[Puthi-Bhandari-Pot]{Communicating Data Narratives: Factuality, Emotionality, and Aesthetics in Rural Bangladeshi Data Curation and Presentation}

\title[Socheton]{\textit{`Socheton'}: A Culturally Appropriate AI Tool to Support Reproductive Well-being}

%the stuff in [] shows up on pages after #1, so it's probably good to make it descriptive / the same

%\author{anonymized}

\author{Sharifa Sultana}
\affiliation{
  \institution{University of Illinois Urbana-Champaign}
  \city{Urbana}
  \state{Illinois}
  \country{USA}}
\email{sharifas@illinois.edu}

\author{Hafsah Mahzabin Chowdhury}
\affiliation{
  \institution{University of Illinois Urbana-Champaign}
  \city{Urbana}
  \state{Illinois}
  \country{USA}}
\email{hafsahc2@illinois.edu}

\author{Zinnat Sultana}
\affiliation{%
  \institution{S.M.R. Law College}
  \country{Jessore, Bangladesh}}
\email{zinnat1409@gmail.com}

\author{Nervo Verdezoto}
\affiliation{%
  \institution{Cardiff University}
  \country{Cardiff, Wales}}
\email{verdezotodiasn@cardiff.ac.uk}

\renewcommand{\shortauthors}{Sultana et al.}

\begin{abstract}
% Modifying abstract
Reproductive well-being education in the Global South is often challenged as many communities perceive many of its contents as misinformation, misconceptions, and language-inappropriate. Our ten-month-long ethnographic study (n=41) investigated the impact of sociocultural landscape, cultural beliefs, and healthcare infrastructure on Bangladeshi people's access to quality reproductive healthcare and set four design goals: combating misinformation, including culturally appropriate language, professionals' accountable moderation, and promoting users' democratic participation. Building on the model of `\textit{Distributive Justice,}' we designed and evaluated \textit{`Socheton,'} a culturally appropriate AI-mediated tool for reproductive well-being that includes healthcare professionals, AI-language teachers, and community members to moderate and run the activity-based platform. Our user study (n=28) revealed that only combating misinformation and language inappropriateness may still leave the community with a conservative mob culture and patronize reproductive care-seeking. This guides well-being HCI design toward being culturally appropriate in the context of reproductive justice with sensitive marginalized communities.
\end{abstract}

%%
%% The code below is generated by the tool at http://dl.acm.org/ccs.cfm.
%% Please copy and paste the code instead of the example below.
%%



% \begin{CCSXML}
% <ccs2012>
%  <concept>
%   <concept_id>10010520.10010553.10010562</concept_id>
%   <concept_desc>Computer systems organization~Embedded systems</concept_desc>
%   <concept_significance>500</concept_significance>
%  </concept>
%  <concept>
%   <concept_id>10010520.10010575.10010755</concept_id>
%   <concept_desc>Computer systems organization~Redundancy</concept_desc>
%   <concept_significance>300</concept_significance>
%  </concept>
%  <concept>
%   <concept_id>10010520.10010553.10010554</concept_id>
%   <concept_desc>Computer systems organization~Robotics</concept_desc>
%   <concept_significance>100</concept_significance>
%  </concept>
%  <concept>
%   <concept_id>10003033.10003083.10003095</concept_id>
%   <concept_desc>Networks~Network reliability</concept_desc>
%   <concept_significance>100</concept_significance>
%  </concept>
% </ccs2012>
% \end{CCSXML}

% \ccsdesc[500]{Computer systems organization~Embedded systems}
% \ccsdesc[300]{Computer systems organization~Redundancy}
% \ccsdesc{Computer systems organization~Robotics}
% \ccsdesc[100]{Networks~Network reliability}

%%
%% Keywords. The author(s) should pick words that accurately describe
%% the work being presented. Separate the keywords with commas.
% \ccsdesc[500]{Online Gender abuse, Social Media, Facebook, Messenger, Solidarity, Feminism,  Bangladesh}


%%
%% This command processes the author and affiliation and title
%% information and builds the first part of the formatted document.

\settopmatter{printfolios=true}

\maketitle

%\newpage
\section{Introduction}
Educating communities about reproductive well-being, including fertility, maternity, genital infections, and sexually transmitted diseases (STDs), has remained one of the major challenges in today's global healthcare sectors. While medical science has advanced with modern knowledge and improved techniques of reproductive health, many of them face challenges in being fruitful while deployed in communities sensitive to certain cultural beliefs and faith-based practices. Such sensitive communities are often suspicious of particular sets of modern reproductive care-related knowledge and find them mismatched to their value systems. Such suspicion leads to misinformation and misconceptions about reproductive well-being. For example, many Bangladeshi people did not agree to contraceptive use initially in the 1970s, as they thought this might be a conspiracy against their religious beliefs that suggest people can only give birth to the number of children that God wishes and humans should not control it \cite{huber1979contraceptive, rahman1980contraceptive}. Additionally, research has found that words, phrases, and humor used in modern reproductive well-being scholarship may not often align with language practices in the community, and therefore, the activities and technologies building on such knowledge failed to serve the community as they were expected \cite{sultana2019parar, sultana2019witchcraft, sultana2020fighting}. For example, many people vulnerable to AIDS refused to gather knowledge and seek help from modern care providers, as several awareness and treatment methods did not use their culturally acceptable sentiments in their contents and practice \cite{54petti2006laboratory, 36truter2007african, nyblade2011once}. 

Along with medical scientists, researchers from the domains of anthropology, social science, and global development have looked into these problems and brought insights into reproductive care-related misinformation and misconceptions. Researchers, activists, and industry have also worked to develop culturally appropriate measures for growing awareness of reproductive well-being. Yet, millions of people across the world are still beyond the scope of reproductive healthcare support. As human-computer interaction (HCI) researchers, we are curious about the depth and breadth of the problems from cultural and socio-infrastructural perspectives, and we look to design possible socio-technical solutions to contribute to this scholarship. 

Existing HCI research on reproductive well-being has developed theories, frameworks, and applications for menstrual care, the sexual health of adolescents, and menopause \cite{55almeida2016looking, kumar2020taking, sondergaard2016periodshare, epstein2017examining, bardzell2019re, campo2019your, svenningsen2020designing, ng2020menstruating}, and advanced the domain with discussions on abortion rights and reproductive data privacy in the West \cite{michie2018her, mehrnezhad2023my}. However, this HCI literature rarely addressed the needs and concerns of communities' sensitive nature toward their value systems. We fill this gap in the literature by engaging with Bangladeshi people who have often shown resistance to many reproductive care technologies and initiatives over time since they did not match their cultural values and faith and/or the language used in them did not match the local acceptable language. 

In our two-phase project, we first conducted an ethnographic study with Bangladeshi men and women (n=41). Through observation, interviews, and focus group discussions, we solicit answers to the following research questions:

\begin{quote}
\textit{RQ1: What challenges do Bangladeshi people face while seeking reproductive care from the formal health infrastructure and the wellness programs available in the country? }\\
\textit{RQ2: Why are modern information and communication technology (ICT)-based reproductive supports challenging to seek care from? }\\
\textit{RQ3: What qualities and affordances would this population like to see in a high-quality and impactful reproductive well-being technology?}
\end{quote}

We found that talking about reproductive care in the community is more stigmatized than we anticipated, as the ethnographer experienced confrontation and verbal abuse in the field. We noted that seeking help for reproductive care from formal healthcare infrastructure is challenging for the participants as those are inadequately resourced, privacy intrusive, and condescending. Additionally, modern ICT-based reproductive support systems are poorly managed and misinformation- and misconception-prone. The participants speculated about affordances in designing future reproductive well-being technologies that will let them actively participate in reproductive care-related knowledge curation. These findings led us to design \textit{`Socheton,'} a culturally appropriate AI tool to support reproductive well-being that allows users and their community to combat misinformation and content with inappropriate languages democratically with professionals and AI moderators.

In the second phase, we conducted a user study with Bangladeshi men and women (n=28) through interviews and focus group discussions. We introduced \textit{`Socheton'} to the participants, discussed the possible benefits this application would bring in supporting their reproductive well-being, and asked them what kind of challenges and troubles this application might generate for the prospective users and other community stakeholders. Our participants were worried that the way today's modern west-centric reproductive care-related knowledge patronizes the space, the scope for mass people's participation may also create a similar scope for radical conservative knowledge and sensitivities in patronizing the space. They also suggested more policy-level intervention to ensure that policymakers and related stakeholders have an accountable role in the process of sharing responsibilities in Distributive Justice for reproductive well-being.  

This work makes four contributions to the human-centric AI design, culturally appropriate AI, well-being HCI, and reproductive justice literature. First, we provide an overview of the existing challenges with formal and informal reproductive healthcare infrastructures that communities sensitive to cultural and faith-based values face in Bangladesh. Second, we designed \textit{`Socheton,'} a culturally sensitive AI tool, to enhance reproductive health care and conducted user evaluation. This tool empowers users and their communities to collaboratively address misinformation and harmful content through democratic engagement with healthcare professionals and AI moderators. Third, our findings from the user study of \textit{`Socheton'} challenge the HCI and AI design's current sensitivities of democratic participation and call for more strategic employment of democracy while developing theories and applications for the community's reproductive well-being. Finally, we discuss the ethical and practical challenges surrounding designing reproductive well-being technologies for Bangladeshi communities sensitive to cultural and faith-based values and similar other communities marginalized in well-being HCI and HCI design scholarship.

\section{Related Work}
The World Health Organization (WHO) defines reproductive health as a state of complete physical, mental, and social well-being and not merely the absence of disease or infirmity in all matters relating to the reproductive system and its functions and processes \cite{rephealth:online}. Reproductive health refers to people's options and capabilities to have a safe sex life and reproduce and the freedom to decide about them \cite{rephealth:online}. Reproductive well-being concerns are associated with people of all ages and genders. Therefore, any violations of reproductive capabilities and freedom are considered ``reproductive injustice" in this paper. %Note: this paper interchangeably uses the terms ``reproductive care" and "reproductive well-being."

\subsection{Reproductive Justice as a Movement}
The history of reproductive (in)justice is old. The first known abortion laws appear in the Code of Hammurabi (Assyria, 1772 B.C.) \cite{reiman1999abortion}. Assyrian women were punished for aborting and allowing fathers to kill newborns, indicating the law controlled women's rights rather than protecting them and the fetus. Ancient Greek and Roman law rarely concerned for the fetus or the mother. An abortion case was only discussed in court if any damage to the husband or his estate was perceived since the woman's body and her unborn children were considered her husband's property \cite{dickison1973abortion}. European colonialism and imperialism profoundly impacted today's reproductive rights and justice worldwide. Colonial powers often imposed their own cultural and medical practices on colonized populations, leading to forced sterilization, trials of medical methods, and limited access to reproductive healthcare \cite{jaspan2005scientific, mugerwa2002first, wailoo2018historical}. For example, American surgeon J. Marion Sims, known as the `father of gynecology,' ran experimental procedures on slave black women without anesthesia and developed theories and surgical methods for modern gynecology \cite{wall2006medical, axelsen1985women}; for the first few decades, the HIV vaccine phase-I trials were conducted in African countries \cite{mugerwa2002first, jaspan2005scientific}, and the initial trials for the birth control pill were conducted in Puerto Rico in the 1950s because of its colonial relationship with the USA \cite{marks2003cage}. 

Being influenced by the industrial revolution, globalization and urbanization, and promotion of population control and eugenics programs, modern reproductive rights movements emerged in the West and the Global South (e.g., South Asian and some Middle Eastern countries) in 1960s and 1970s, advocating for women's right to control their bodies and reproductive choices \cite{chhachhi1989state, correa1994population, bracke2022women, ross2017reproductive}. The United Nations (UN) Officially acknowledged the reproductive (in)justice agendas at the International Conference on Population and Development (ICPD) in Cairo in 1994, and today is an advocate for global reproductive justice \cite{el1999political}. Today's reproductive justice agendas include capabilities of abortion; reproductive healthcare disparities across race, gender, ethnicity, and religion; reproductive justice and climate change; and LGBTQ+ equality \cite{de2012looking, hessini2007abortion, price2018queering, arousell2016culture}.

\subsection{Reproductive Well-being in HCI}
Reproductive well-being has a long chain of literature in HCI and social computing. For example, Almeida et al.'s critical analysis of theories and frameworks for designing intimate wearables and designing tools to overcome the stigma around testing for pelvic fitness \cite{55almeida2016looking, almeida2015designing, almeida2020woman}. Researchers have developed theories and designed for menstrual care, the sexual health of adolescents, and menopause \cite{sondergaard2016periodshare, epstein2017examining, wood2017sex, mcdonald2018maxifab, lazar2019parting, bardzell2019re, campo2019your, campo2020touching, sondergaard2020designing, svenningsen2020designing, ng2020menstruating}. Pressing concerns such as abortion rights and reproductive data privacy were also discussed in HCI scholarship \cite{michie2018her, mehrnezhad2023my}. While this literature has significantly advanced the domain, many of their findings suggest that the knowledge and implications produced by them are socially and culturally constructed, context-dependent, and may not apply to other cultures in the Global South. %Therefore, our research borrows motivation from these projects while drawing primarily on reproductive care literature in the Global South and Bangladesh, which we discuss in the following subsections. Medical mistrust from a mismatch between Western medical practices and local cultural beliefs frequently leads many medical technologies and procedures to fail in these communities.

Global South reproductive well-being scholarship has primarily concentrated on the troubles of mismatching values between cultural norms and modern medicine and technology design. For example, certain African communities believed that foreign substances could harm their bodies that are intrinsically connected to nature, and therefore rejected HIV vaccinations \cite{54petti2006laboratory, 36truter2007african, nyblade2011once}. Similarly, contraceptives and many pregnancy-care methods also faced significant resistance in India, Pakistan, Kenya, and Egypt \cite{ghule2015barriers,eshak2020myths,declerque1986rumor,ataullahjan2019family,casterline2001obstacles,akoth2021prevalence,britton2021women}. While designing ICT-based support for reproductive well-being of Kenyan women, Perrier et al. argued for involving humans in the loop to address context-sensitivity \cite{perrier2015engaging}. Research with Arab Muslim community suggested considering cultural and religious aspects of women's intimate health in such designs \cite{al2024understanding}. Research on Indian women's maternal health, young adults' learning processes, knowledge discrepancies, and information exchange behavior on reproductive well-being also advanced the domain \cite{kumar2015mobile, kumar2015projecting, tuli2019sa, tuli2022rethinking, tuli2018learning, kumar2020taking, tuli2020menstrual, jain2015game}. For example, Bagalkot et al. show how the embodied pregnancy experiences of Indian women are influenced and negotiated by the socio-cultural context and existing care infrastructure, often through conflicting norms, beliefs, and practices of medicine, nourishment, and care \cite{bagalkot2022embodied}. In addition, Mustafa et al. informed the domain about Pakistani women's prevalent maternal health beliefs and religious practices influential to reproductive care decision-making \cite{mustafa2020patriarchy, mustafa2021religion}. Sultana et al. also found a significant influence of cultural and spiritual norms in Bangladeshi women's healthcare decision-making, especially around pregnancy \cite{sultana2020parareligious, sultana2019parar, sultana2019witchcraft, sultana2020fighting}. 

Noteworthy that Bangladesh holds a complex blend of many different values, including local traditional norms, religious sentiments, spirituality and faith systems, patriarchy, lately introduced liberal values, etc., all of which significantly influence women's autonomy over their reproductive choices \cite{sultana2023computing, sultana2018design}. Through different government and non-government organization (NGO) initiatives, maternal morbidity rates and family planning practices have improved \cite{dhsprogr7:online, koenig2007maternal, joshi2013family}. However, introducing these techniques in faith-sensitive communities faced significant backlash \cite{rahman2016mayer}. To date, Bangladeshi adolescents often lack comprehensive knowledge of reproductive health due to limited access to and stigma of age-appropriate sex education, which contributes to increasing the risk of teenage pregnancy and infection of sexually transmitted diseases \cite{sayem2011factors, sarder2020determinants, hossain2017knowledge}.

However, technology-mediated well-being support systems are challenged and fail in Bangladeshi communities because of cultural resistance, mismatching values, low infrastructural support, and people's disbelief in the whole ecosystem \cite{sultana2020parareligious, sultana2019parar, sultana2019witchcraft, sultana2020fighting, sultana2021dissemination}. Yet, services like Aponjon provide an SMS-based digital healthcare service for pregnant women, new mothers, and families \cite{Aponjon14:online}. Additionally, researchers have developed and tested many different tools and techniques (e.g., Pregnancy Tracker, GorbhoKotha, SmartCare, etc.) in lab settings to support maternal care in Bangladesh \cite{mahbub2024gorbhokotha, tumpa2017smart, kundu2020evaluating}. This scholarship suggests that designing for Bangladeshi reproductive well-being requires growing extensive knowledge of the community's cultural sensitivity and involving appropriate stakeholders, which we address through this research.

\subsection{Community-Centric Design of Reproductive Well-being}
Building on existing literature on reproductive justice, we strategize our reproductive support to be community-centric. Through this support, we intend reformation, including education, awareness, collaboration, and direct interventions. At the same time, the design would consider the potential concerns while aiming to combat the stigma of reproductive awareness through distributed responsibilities. Such concerns might include resource constraints, an unfair extra workload, additional stress, etc. Therefore, our design motivation also aligns with the idea of \textit{Distributive Justice}, where the welfare-based principles suggest distributing the material goods and services in the society in a way that contributes to social welfare and further allows the community members to share the welfare responsibilities equitably \cite{schmidtz1998social, schroth2008distributive, sen1982utilitarianism, Rescher66dis}. \newedits{Welfare-based principles of distributive justice prioritize maximizing overall social well-being by allocating resources in ways that improve collective welfare, rather than strictly adhering to individual rights or desert \cite{schroth2008distributive, sen1982utilitarianism, murphy2002myth}. These principles also emphasize the fair distribution of responsibilities, encouraging community members to contribute to and share in sustaining societal welfare.} Some recent HCI-design works on gender justice and child sexual abuse in the Global South have built on this model of justice and received impactful feedback \cite{sultana2021unmochon, sultana2022shishushurokkha, chordia2024social, coleman2023reconsidering}. Hence, we build on the model of \textit{Distributive Justice} to investigate our research questions and address them through design.

%\subsection{Culturally Appropriate and Value Sensitive Language Practice in HCI and AI}

%\newpage
\section{Phase-1: Understanding Reproductive Well-being in Bangladesh}
Phase 1 solicits answers to RQ1, RQ2, and RQ3 through an ethnographic study and sets appropriate design goals based on the findings. Below, we discuss the methods and findings and the process of translating them into design decisions. 

\subsection{Methods}
\begin{table}[!t]
%\begin{wraptable}{r}
\centering
\begin{tabular}{|rl|}
\hline
Total Participants: & 41 (Female: 24, Male: 17)\\
\hdashline
\multicolumn{2}{|c|}{\textbf{Type of Participation}} \\
\hdashline
Interview only: & 8 (Female: 5, Male: 3)\\
FGD only: & 15 (Female: 8, Male: 7)\\
Interview + FGD: & 18 (Female: 11, Male: 7)\\
\hdashline

\multicolumn{2}{|c|}{\textbf{Age range (in Years)}} \\
\hdashline
All: & 19-60, median 33\\
Female: & 19-60, median 30\\
Male: & 19-56, median 38\\
\hdashline

\multicolumn{2}{|c|}{\textbf{Education}} \\
\hdashline
No Formal Schooling: & 15 (Female: 8, Male: 7)\\
Primary School: & 18 (Female: 10, Male: 8)\\
Secondary School: & 5 (Female: 3, Male: 2)\\
Higher Secondary School: & 2 (Female: 2, Male: 0)\\
Undergraduate and Above: & 1 (Female: 1, Male: 0)\\
\hline
\end{tabular}
\caption{Demographics of the participants in interviews and focus group discussions (FGDs)}
%\vspace{-25pt}
%\end{wraptable}
\end{table}

In the first phase, a ten-month ethnographic study was conducted between October 2021 and January 2023 in three villages located within a 10-15 kilometer radius of Jessore town, Bangladesh. The study sites included Rodropur, Chachra, and Shankorpur. Jessore district has historically been a focal point of development in Bangladesh, with a recent surge in healthcare infrastructure. Over 55 private and public hospitals have been established in the district, many specializing in reproductive care. Additionally, more than 100 NGOs operate in the region, with over 30 providing programs focused on women's health, including pregnancy, childbirth, and menstrual care. The ethnographer was born and raised in Jessore and is familiar with many local cultural practices. In those ten months, we studied with 41 interview and focus group discussion participants. The fieldwork consisted of semi-structured interviews (n=26), focus group discussions (n=33), and observational field notes of rural healthcare and well-being accompanied by contextual inquiries and photography.

\subsubsection{Access to the participants}
Our access to participants in the villages was facilitated by the Rural Reconstruction Foundation (RRF), a non-profit global development organization that offers microfinance, education, health, and agricultural programs \cite{34RRF}. Their officials introduced us to front-line microcredit workers who visit rural clients' homes weekly. RRF workers helped the ethnographer reach such communities by taking her to villages where they worked. After arriving in the village, the ethnographer held a public community meeting with the microcredit clients and explained the purpose of the research study. After answering the microcredit clients' questions and concerns, we recruited participants from the meetings based on their availability. We were interested in engaging with both men and women in the villages and solicit about their reproductive healthcare and wellbeing practices. Further recruitment was performed through snowball sampling \cite{Biernacki1981}.

The primary occupations of those families include farming, fishing, and small businesses, with an average (and median) monthly income of approximately USD 100. We developed a close relationship with those families by making frequent visits, engaging in long conversations, helping with their household and daily activities, and joining their afternoon hangouts while observing them in a participatory fashion. After building rapport this way, we slowly started asking them about our queries about their reproductive healthcare and well-being. In addition to the performers and group managers, local villagers also participated in these meetings and shared their feedback. While joining as observers, we recruited many participants from such meetings. 

The ethnographer is a native Bengali speaker and has a long-term familiarity with the neighborhood. She was born and raised in Jessore and has lived in various parts of the district. This positionality helped her access the population and build rapport with participants. All interactions with participants were conducted in Bengali, the local language, which all the participants and the ethnographer speak fluently. We obtained oral consent from the participants in consultation with a university institutional review board since many villagers were low-literacy and would have trouble reading and understanding a written informed consent form.

\subsubsection{Observation}
We started our study by observing our participants to understand their daily lifestyles at home and in social settings. We observed their daily work, hangouts, regular meetings with other friends, and enjoyable activities, focusing on understanding how they take care of their personal and family's physical and reproductive well-being. During the observations, we asked them situated and spontaneous questions to understand their activities better. The mode of the observation was participatory. We participated in each of these activities with the participants' permission, observed their activities and responses, and recorded them in our notes. We conducted around ten hours of observation, spending three to five hours in each village.

\subsubsection{Focus Group Discussion (FGD)}
We also conducted focus group discussions with 33 participants in the villages. Each group consisted of four to eight participants. The focus group participants were identified through snowball sampling with the help of RRF microcredit fieldworkers. Discussion topics included their daily lives, the kinds of social, physical, affective, and reproductive issues they suffer, how norms and cultural practices were influential in their community, their experiences and treatment by their families while sick, and how they felt about their local reproductive well-being support systems. We also investigated their current access to and use of technology (such as mobile phones, computers, and the internet, etc.). We asked about their social inclusiveness and support, including potential frustrations, stigma, rumors, and superstitions around reproductive care, how they worked around them, and how they imagined these concerns were addressed in a hypothetical ICT tool to support their reproductive well-being. The sessions were generally 35 to 45 minutes long. We took detailed notes and audio-recorded all the discussions with the participants' permission.

\subsubsection{One-on-one Interviews}
We conducted one-on-one interviews with 26 participants. Each interview lasted approximately thirty minutes and was conducted wherever it was convenient for the participant, often in their homes. Participants were sought based on rapport with the ethnographer in earlier sessions and further through snowball sampling. Discussion topics included their daily lives, the kinds of social, physical, affective, and reproductive issues they suffer, how norms and cultural practices were influential in their community, their experiences and treatment by their families while sick, and how they felt about their local reproductive well-being support systems. We were also interested in their collaborative practices within the community regarding such stigmatized issues. Additionally, we asked them about the challenges they face in the modern day, how they resolve them, how many of the solutions involve computer-mediated systems and social media, and how they imagined these concerns were addressed in a hypothetical ICT tool to support their reproductive well-being. Interviews were again semi-structured, so we asked further related questions to understand the participant's responses and go deeper into topics. Each interview lasted around 30-40 minutes. We took detailed notes of all the interviews and audio-recorded ten (avg. 25 mins) with participant permission.

\subsubsection{Data Collection and Analysis}
We collected approximately three hours of audio recordings and 110 pages of field notes, which were transcribed and translated into English. We then performed thematic analysis on our transcription \cite{76boyatzis1998transforming, 75strauss1990open}, starting by reading through the transcripts carefully, allowing codes to develop. Twenty-three codes spontaneously developed initially. After a few iterations, we clustered related codes into themes: menstruation, menopause, black blood, childbirth, stillborn, mid-wife, stigma, materials, myth, religion, spirituality, etc. Our major findings were categorized based on them.

\subsection{Unpacking Aspects of Reproductive Well-being in Bangladesh}
This section dives into how villagers in our study approach their reproductive well-being and the support systems available to them with associated challenges, how accessible and useful ICT is in these communities, and how technology could be designed to improve their access to quality reproductive healthcare.

\subsubsection{Available Reproductive Well-being Resources in the Fieldsite}
Our participants told us that they seek help for their reproductive well-being from both formal and informal care-providing infrastructure. In villagers' definition, the \textbf{formal healthcare-providing infrastructure} follows modern medical healthcare services and is run on government approvals. Jessore has one general (or, \textit{`Jessore Sadar'}), ten other specialized hospitals, twenty-five private clinics, eight health centers, and 22 rural sub-centers, all government-subsidized. More than 300 doctors serve in Jessore, with available doctor-to-patient ratios of 1:13812 \cite{khulna:online, Jessore1:online}. Sub-centers are run by a small medical team that handles first-aid, minor stitching, normal deliveries, and some non-complicated illnesses treated with oral medication. Generally, they do not conduct surgery, but natural child delivery is operated in such centers. Occasionally, pediatricians, urologists, and gynecologists of other larger hospitals serve these centers. %Our participants are also used to seeking help from private clinics. 

Additionally, non-government organizations (NGOs), including RRF, the HOPE Foundation for Women \& Children of Bangladesh, the White Ribbon Alliance Bangladesh, Enfants du Monde, BRAC, CARE Bangladesh, Marie Stopes Bangladesh, and others, run field-level reproductive well-being programs. NGOs often establish temporary field offices and recruit locals as data collectors, communicators, and educators. The workshops typically incorporate educational sessions and health screenings, focusing on hygiene, maternal health, vaccinations, and family planning. One village reported a program on adolescent mental health with negligible attendance. %For longer-term programs, NGOs actively involve the community in data collection and conduct workshops on childcare and family planning.

Following villagers' definition, the \textbf{informal care-providing infrastructure} does not adhere strictly to modern medical practices but often enjoys government approval and widespread popularity within communities. We found that rural witchcraft practices, traditional healers, herbal medicine practitioners, midwives, and village market medicine shops are highly accepted information care sources in rural communities. Villagers reported a longstanding tradition of regarding witchcraft as a means of spiritual well-being, dating back hundreds of years. Numerous tales and local myths describe the influence of supernatural entities, such as Jinn, Voot, and evil spirits, on villagers' lives, local politics, and community development through their impact on wealth and childbirth. Witchcraft practices often involve traditional herbal medicine methods and materials (plant and soil-based substances) to address the reproductive health of their clients. Midwives and shopkeepers of medicine shops occasionally have modern medical training with limited scope, yet they are highly accepted care providers in rural communities.

\subsubsection{Constraints with Conceptualizing, Decision Making, and Resources}
The neighborhoods we studied are predominantly conservative Muslim and highly patriarchal. We found that the process of being introduced to reproductive well-being was gendered in the communities. All the women learned about personal care, menstruation, pregnancy care, and childcare from their mothers and other female elders. However, most male participants learned from mosques, friends, and internet contents. Ten male and five female participants told us that fathers and brothers in the family are traditionally reluctant to discuss such things; instead, they often request the mosque Imam (leader of the prayer) to discuss these things after prayer. Three groups of FGD participants told us about the government's inclusion of menstruation and adolescence in the high school curriculum. However, the participants were worried that not everyone might embrace this change, as at least ten told us that their children's schools skipped those chapters in the class. We also noted two stories where NGOs tried to help schools by setting up separate sessions for boys and girls at the schools, but weird rumors spread out; they aborted the sessions and apologized to the communities. Discussion on reproductive well-being topics is highly stigmatized and generally forbidden if it is not with closed ones. For example, two non-participant women dragged their sister and daughter-in-law out of the sessions during two different focus group discussions and confronted the ethnographer. One of them said, 

\begin{quote}
\textit{``You girl better keep your shameless American monkey business out of our area in broad daylight. Do not try to make our girls brazen. They are respectful women; they keep their "Haya" (things matter of shame or to be protected) behind purdah.", (fieldnote, village 2)}
\end{quote} 

More than 30 participants told us they relied on \textbf{self-diagnosis} and \textbf{domestic consultancy}. Twenty-two male and female participants revealed that awareness of potential reproductive illness often arose from experiencing physically noticeable and measurable symptoms or seeing someone suffering. However, seeking professional help for women was less important to their families, as ten female participants reported that their families asked them to prioritize their heavy workloads at home above their own well-being, attempting natural remedies and keeping their illnesses private unless they were pregnant with a male child. Seven male participants reported ignoring the symptoms based on domestic consultancy until their conjugal life and mobility were hampered. 

We also found that many patriarchal and conservative practices influenced the participants' reproductive help-seeking. The female participants were restricted from visiting hospitals or doctors independently. Twelve women said they needed their husbands', or male family heads' permission first. They also added that permission-seeking is often challenging if the male family heads struggle to understand their private needs. Ten female and five male participants told us that women often struggle to explain their reproductive health issues because they lack vocabulary for phenomena and experiences as those discussions are stigmatized, so they believed a senior companion from home would be handy in such cases. 

The practitioner's gender was also a concern to many participants. Fifteen female participants said they would be shy to answer a male gynecologist or other doctors if asked questions about their reproductive well-being. They disapproved of being touched by male doctors in those areas during medical tests. Similarly, fourteen male participants also struggled with female doctors. Our female participants reminded us about the patriarchal nature of the society where female professionals' skills and experiences, even if they are highly qualified, are disregarded because of their gender. Some male participants also thought female doctors do not know enough to understand men's problems; a participant said, 

\begin{quote}
\textit{``My wife's female gynecologist asked me private questions about my pelvic areas and stuff. It was shameful for a woman to ask a man. She is not a man; she does not understand men's issues, and I did not bother answering her properly. It is humiliating.", (P29, Male, Interview)}
\end{quote}

At least seven other male participants were similarly skeptical of female doctors' qualifications in understanding men's problems and found thought conversations humiliating because of a lack of culture and value sensitivities. 

However, informal healthcare is still prioritized for pregnancy by many participants' families for being culture and value-sensitive. By tradition, pregnant ladies are sent to their parent's houses for childbirth regardless of whether the neighborhood has proper hospitals. In many cases, her family would bring midwives for childbirth. Also, midwives are the top choice to operate abortions secretly if the child in the womb is a daughter, as aborting because of the child's sex is illegal in the country. Ten women told us stories of them or someone in their families where such secretly arranged abortions at home severely damaged their bodies as midwives lacked skills. We quote one, 

\begin{quote}
\textit{``Several years back, when I conceived and found out that it was a girl, my mother-in-law was furious and wanted a divorce between me and my husband, as my firstborn was also a girl. I wanted to keep the child, but my parents said they would not take responsibility for a divorced daughter and her daughter. I begged on my knees; even the doctors said it was too late to abort the child. They arranged for a midwife and bought some medicines to abort the child at home. The child was aborted, and my organs did not heal well. I am having trouble getting pregnant again, and the whole family regrets it now.", (P17, Female, Interview)}
\end{quote}

While we noted two maternal deaths and two stillbirths under midwife supervision, still seven female and three male participants prioritized local informal support over hospitals because of low resources and privacy concerns. All the participants told us that they could barely have proper conversations with the practitioners in crowded hospitals, let alone seek further explanations or argue, as most doctors let multiple patients enter their chambers so that they could finish up seeing hundreds of patients by the end of their work hours. At least twenty-five participants in multiple focus group discussions expressed their frustration over this, as one woman mentioned,

\begin{quote}
\textit{``We went to see Dr. Nazmun (pseudo name), the best gynecologist in the town. Our turn came at 11 pm late at night. We entered her office and found three other patients and their companions waiting; one of them was a patient's husband. We needed her to check the private parts of my daughter, so we waited until the man left, but the doctor's assistant yelled at us for procrastinating while the doctor was still talking to two other patients simultaneously. So I yelled back because my daughter deserved her privacy. I did not take my daughter to a female doctor and waited until 11 pm to get her undressed before a man.", (P4, Female, FGD-1)}
\end{quote}

Thus, the villagers explained their challenges of resource constraints and additional byproducts and how they often needed to choose between their time, privacy, and scope of receiving quality reproductive care.
%%%% insert about ICT

\subsubsection{Mismatch and Mistrust in Language, Values, and Methods}
More than twenty participants reported about practitioners' unfamiliar language. For instance, seven female and five male participants mentioned that they struggled to understand the medical terms the doctor used to explain their physical conditions. Sixteen participants expressed that their traditional views and practices clashed with the healthcare provided by the complex and were not always accommodated with empathy; as one of them told us,

\begin{quote}
\textit{``We take `Kabiraji' (religion- and spirituality-based traditional healing system) and herbal medicines from the local traditional healers. Anything described in `Kabiraji' language is very understandable. However, doctors hate to listen to those words and interrupt us. They have shut me down many times because they did not understand me or hated to hear me out.", (P9, Female, FGD-2)}
\end{quote}

Such mismatching language was troublesome for young adults, as participants and practitioners shared stories from experiences. A participant was a village doctor's assistant who told us that most young girls and boys in their teens are very shy about the changes in their bodies and are often misguided by people around them. Thus, they do not often come across proper scientific and medical terms associated with their reproductive well-being, and due to stigma, they do not get to learn as they grow further. The participant told us a case of a girl whom they recently gave consultancy to, as we quote her, 

\begin{quote}
\textit{``A 13-14 years old girl sought help from her neighborhood sister for the acne around her vaginal areas. That sister referred to bad spirits and advised her to apply weird stuff. The girl got an infection. She skipped school to travel miles afar to avoid known people and see us. However, she spoke jargon rooted in the genre of Jinn magic, black magic, and local spirituality scholarship that the doctor did not understand. I know those languages, so I had to interpret to help them communicate.", (P14, Female, FGD-3)}
\end{quote}

However, the problems with mismatching language were bidirectional, as the many phrases and sentiments used in modern West-centric reproductive well-being contents were unacceptable to them in public and sometimes in private discussions. \label{lgbtq-book}For example, a group of FGD participants discussed an event on national news and national TV in January 2024, where two university teachers led a social media mob against including LGBTQ concepts in the national high school curriculum (refer to \cite{Sharifa68:online} for context). They said, 

\begin{quote}
\textit{``P38: The chapter talked about a person, "Sharif", who was raised as a man. But soon, he found out about other people out there, just like him --- not exactly men. So he goes to be "Sharifa" (a woman).\\
P35: The professors said the chapters would influence the children from converting "Sharif to Sharifa", like from man to woman. This is absolutely against Islam; why would someone change their sex and gender from something they were born with? It was perfect that they tore off the book chapter and burned it. \\
P37: But Sharif was born Hijra (someone with chromosomes other than XX and XY), so he decided to convert. \\
P35: That is the problem; sex change operation is haram, and promoting it in children's books is also disgusting. I will never accept if my son Karim decides to be Karima tomorrow.", (Male, FGD-5)}
\end{quote}

Two other focus groups discussed this January'24 high school book-burning event and emphasized that reproductive well-being contents should align with the cultural sentiments of the community. While the participants confirmed that they support Hijra's welfare, they also argued for the proper choice of concepts, sentiments, and phrases in such educational materials.  

We also found some of the reproductive well-being methods suggested by the local doctors that they found conflicting with their practiced sentiments. For example, a group of FGD participants expressed disappointment when a local health worker suggested using menstrual cups instead of homemade sanitary cloth napkins. We quote one of them, 

\begin{quote}
\textit{``She said that cup was more hygienic and reusable. As soon as she said it had to enter through the vagina, that was a major turn-off. Nothing should pass through that channel of an unmarried girl, or that will be a sin. The whole family will be just done, if anyone knows; they won't be able to marry her off! \\
: Well then, men need to learn about cups' benefits first so that they cause no trouble when they find out that their wives have been using them." (P11, Female, FGD-3)}
\end{quote}

The participants acknowledged that cups are good products if used with the family's acceptance and understanding and emphasized that promotions of such products should follow cultural sensitivities and find appropriate targets.

\subsubsection{Community Learning, Accountable Moderation, and Authenticity}
Along with in-person peer learning, all the participants told us they had heard of online-based reproductive well-being support platforms. Twenty-seven participants told us they often find videos on this topic on their YouTube scrolls. However, several struggled with saving those videos for later. Another 21 mentioned they had heard of Reddit sub-groups, Quora groups, and different Bengali and English healthcare and well-being blogs. However, all of them told us that, in most cases, they were unsure how to verify new and suspicious information, as they never received responses upon contacting their moderators. Twelve participants also blamed the language barrier in such cases. 

Seventeen participants were active members of some Bangladeshi Facebook groups where people consult about well-being issues and share knowledge from their experiences with doctors or medical treatments. The participants said members often ask about recommendations on whether to see a doctor or use home remedies, which doctor to seek, and reviews on doctors and hospitals, among others. However, all the participants in this subset told us stories of Facebook misjudging their concerns and frequently troubling their posts, reviews, and queries in the name of content moderation. We quote one of the participants, 

\begin{quote}
\textit{``We have a Facebook group where people generally post about that kind of problem. A group member once posted about their erectile-related problem using local jargon, "Blackwater/Kalpana." However, some members found it inappropriate, and people discussed about allowing this post. Then, the group admin suddenly told us that Facebook had asked them to delete that post as it did not fit the Facebook community standards. Otherwise, Facebook might even close down the group. The admin had to delete it, and the group members were upset.", (P40, Male, FGD-6)}
\end{quote}

Thus, the community found Facebook's forced moderation policy to be irresponsible and insensitive. Eighteen female and ten male participants opined that people's experiences shared through stories help them understand the contexts and circumstances better and learn. However, the participants also discussed challenges associated with false and misleading claims on social media during the discussions of a possible design of ICT-based reproductive well-being support, 

\begin{quote}
\textit{``Not everyone's story on social media is true. But if there is someone we can trust to be a good evaluator of medical facts, at least, I think that will help mass people establish faith in the contents. Maybe the tool can allow us to challenge, and the evaluator can give us a second opinion with more sources and feedback.", (P37, Male, FGD-5)}
\end{quote}

The design discussion part of this and other FGD and interview sessions brought up the challenge of the authenticity of information and responsible content moderation, along with culture and value-sensitivities, on the hypothetical ICT-based support system for their reproductive well-being.

\subsection{Summary and What is missing}
In summary, we found four major aspects of Bangladeshi reproductive well-being and related challenges. First, participants informed us of the lack of ways to freely ask, learn, challenge, and verify their confusion about these sensitive and stigmatized topics. Second, even when they found materials, they could not find people accountable for conveying their concerns about inappropriate language use. Third, in many cases, the participants believed that if the contents had been curated with assistance from the community, it would have had more impact. In fact, they wanted a reflection of their voice in the reproductive well-being content for and by them. Fourth, participants informed us that most Bangladeshi reproductive well-being initiatives involve women's concerns and ignore men's concerns, while people of all genders are vulnerable. Therefore, they suggested a more community-centric approach over women-centric approaches. 

To our knowledge, no active application or platform in Bangladesh currently offers such affordances to its citizens. We pick the challenge of designing a community-centric reproductive well-being support system. We did not prioritize this objective over the other design needs, but to complement them.  
%\newpage
\section{Designing \textit{`Socheton'}: An AI-mediated Tool Supporting Reproductive Well-being}

We drew on the findings from phase-1 and translated those into designing \textit{`Socheton'}, an AI-mediated tool to support reproductive well-being for communities sensitive to cultural values and faith-based practices. It is a Bengali adjective meaning `conscious', `concerned', and `aware.' A group of pre-design participants discussed a hypothetical tool named `Socheton' to support reproductive well-being in the community. Later, we borrowed it to name our tool. 
%This section describes our design goals and components of the application, followed by details on the implementation. 

\begin{figure*}[!t]
\centering
    \includegraphics[width=0.9\textwidth]{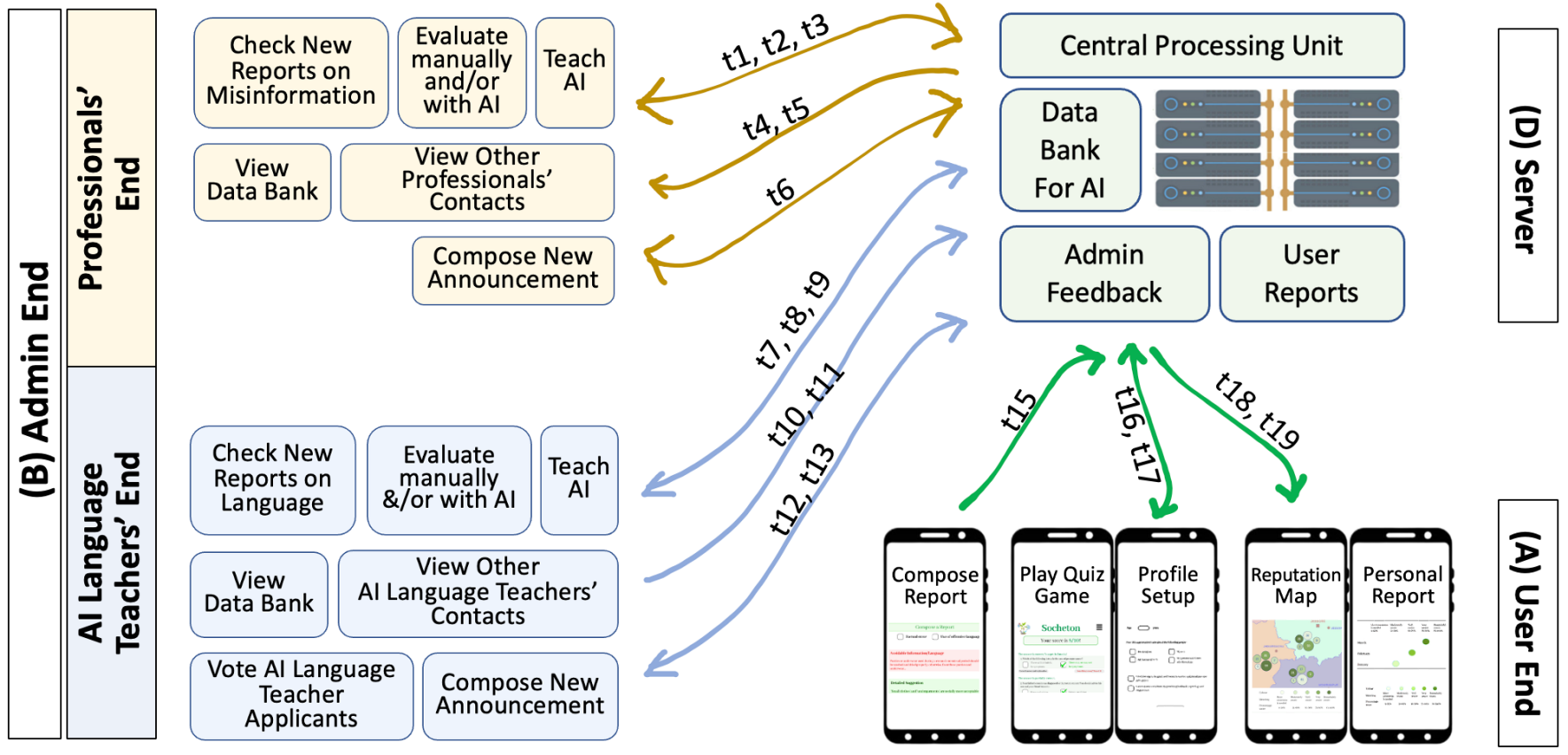}
    \caption{Work-flow diagram of \textit{Socheton}. (A) User End lets users set up profiles, play quizzes, check personal records, compose reports, and view neighborhood reputation maps. (B) Admin End lets health professionals review suspicious information and send them to the data bank in the server for AI to learn, view the data bank and other admins' contacts and compose new announcements; and let AI-LT review reports on language, send the decisions to data bank for AI to learn, compose new announcements, and communicate with other admins; (D) Server holds admin feedback, users' reports, data bank and functionalize the (C) application including AI. It connects to the Professional's, AI-LT, and User end via t1-t6, t7-t13, and t14-t19 tasks to run the application.}
    \label{fig:figure1}
\end{figure*}

\subsection{Design Goals}
In phase 1, we found that the participants struggled to verify misinformation and misconceptions, were concerned with the inappropriate language used in modern Western knowledge of reproductive well-being content, and restricted access to reproductive well-being content curation. Therefore, we set our objective to help Bangladeshi communities collaborate with professionals and curate culturally appropriate content for reproductive well-being. We thought of including AI to allow professionals to manage mass communication requests from many users more easily. We draw on \textit{`Distributive Justice'} for sharing loads in community-based design approaches used in HCI design \cite{sultana2021unmochon, sultana2022shishushurokkha} and shame-based design in HCI design that uses shame as a drive to motivate the community to establish justice for gender and sexual harassment and abuse \cite{blackwell2017classification, sultana2021unmochon}. Our design goals were: 

\begin{quote}
\textit{\textbf{G1: Combating Misinformation and Misconceptions.}} The participants described scenarios where they encountered their perceived misinformation and misconceptions online and in the real world and struggled to find ways to verify them. This kind of scenario urges us to set the goal of designing a tool that helps them report it and get feedback.

\textit{\textbf{G2: Culturally Appropriate Language and Presentation.}}
Several participants pointed out that much awareness content on reproductive well-being and sex education uses language that is culturally incoherent and inappropriate in their views, and they wanted a way to collaborate with the content creators for culturally appropriate content. This scenario urges us to set the goal of designing a tool that reports inappropriate language (e.g. word, presentation, humor, etc.) and suggest possible alternative language. This will also make sure their voices are heard.

\textit{\textbf{G3: Accountable Moderation.}}
Participants often found content that refers to some entities that are hard to communicate and make accountable. This urged us to design a tool where the reference of knowledge or its verifiers could be approached and asked for feedback on participants' reports in practicing accountable moderation. 

\textit{\textbf{G4: Democratic Participation.}}
Participants wanted their stigma to be considered with empathy and their voices to be heard and considered with care. This urged us to design a tool that will channel their voice to content curators and allow them to get accountable feedback.  
\end{quote}

%%  add to feedback: and if rejected, explain with rationales

\subsection{Components and Workflow of \textit{Socheton}}
%\subsubsection{Design Components}
\textit{Socheton} has four components: (A) Users, (b) Admins, (C) the Application (with AI assistance), and (D) the Server. Refer to Fig.1 to follow their details below:

\textit{(A) Users. }
Users are expected to be the community members who will use the app. Upon opening the app, they will set up their profile with age, gender, and choice of content based on the state of their family members and themselves. They can play quiz-based games, challenge the contents and explanations, and report and send feedback on content if they suspect them of misinformation and inappropriate language (refer to Fig.3). They can track their progress from personal records of quiz scores and their community's standing among other neighboring areas. 

\textit{(B) Admins. }
``The admins" consist of professionals and AI-language-teachers (AI-LT, henceforth). The professionals include local doctors, healthcare workers of local influential NGOs, and concerned members with extensive knowledge of reproductive well-being in the community. They can post new content, scrutinize users' reports of misinformation, and command AI to learn from their evaluation (Fig.2 (a2-d2)). The AI-LT includes people with knowledge of both local language and application management. They will scrutinize users' reports of inappropriate language, fix them with appropriate and socially acceptable language \newedits{(words and phrases that are rooted in the local culture and do not have obscene expressions and interpretations.)}, and command AI to learn from their evaluation (Fig.2 (a3-d3)). They will also vote for applicants to recruit for AI-LT positions. The community will set initial criteria and qualifications for prospective candidates and elect admins from the candidate pool.  

\begin{figure*}[!t]
\centering
    \includegraphics[width=0.99\textwidth]{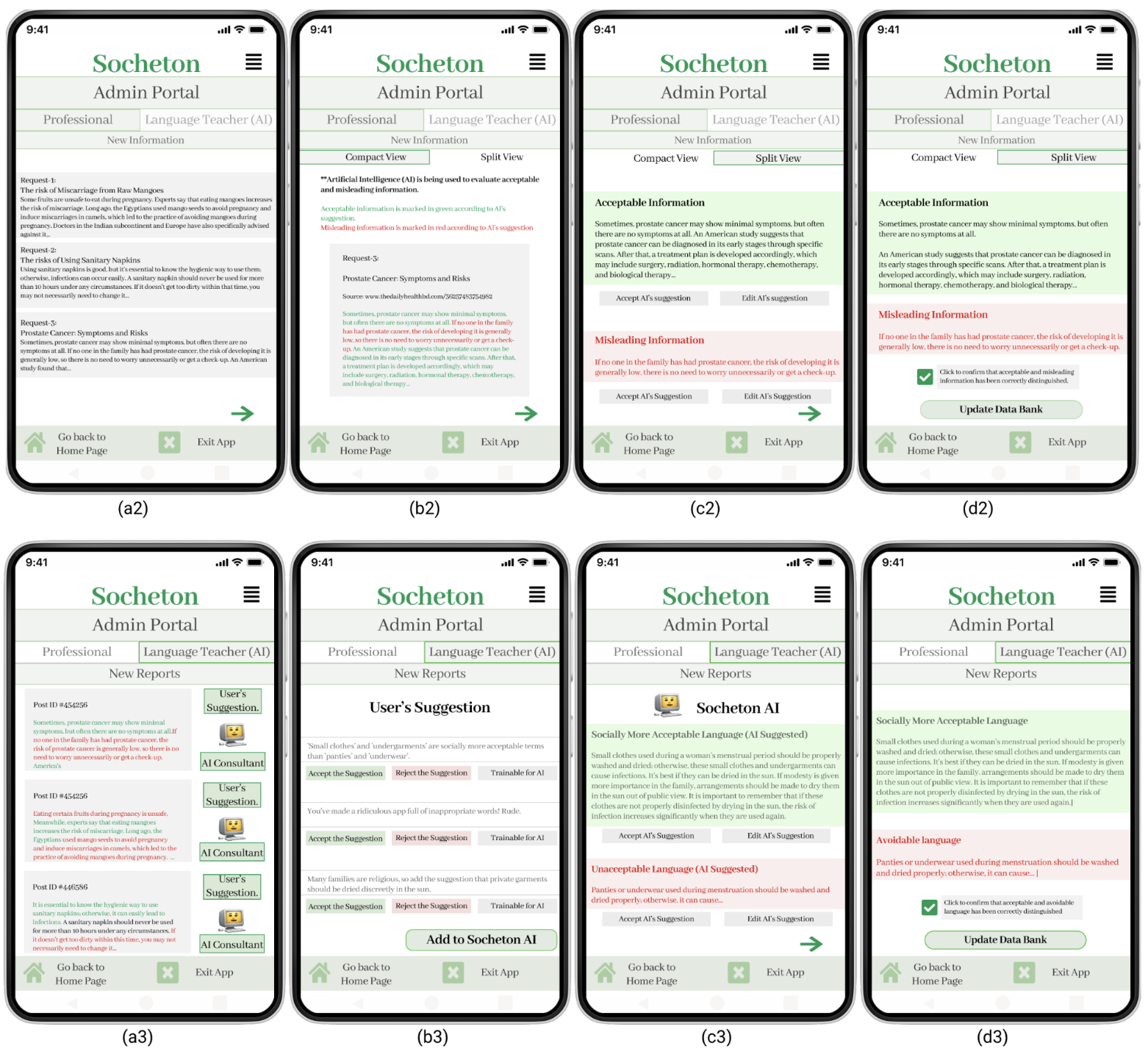}
    \caption{Example pages from \textit{Socheton} Application's Admin End. (a2-d2) Professionals' POV of log of new reports on misinformation from users, a compact and a split view that highlights right and wrong information suggested by AI, allows them to edit and send to data bank for AI to learn for future use. (a3-d3) AI-TLs' POV of log of new reports on language from users, users' and AI's suggestions, console for them to edit and send to data bank for AI to learn for future use.}
    \label{fig:figure1}
\end{figure*}

\textit{(C) Application. }
The application part of \textit{Socheton} is an Android app for admin and user groups with AI assistance. It offers the professional admins the options: (ai) view new reports, (aii) view information bank, (aiii) view/contact other professional admins, and (aiv) compose new posts to functionalize G1, G3, and G4. The first page of the view report contains the log of all the new reports along with the view option of AI suggestions and users' suggestions on misinformation. Upon clicking one report, the professional admin can watch detailed suggestions by users and AI on separate pages and decide to accept, reject, and edit them before commanding AI to learn them for future suggestions. The professional admin can view/contact other admins and compose new content with references. They can also view the information bank used by AI.

To functionalize G2, G3, and G4, \textit{Socheton} offers the AI-LT admins the following options: (bi) view new reports, (bii) view/contact other professional admins, and (biii) vote for recruiting new AI-LT. The first page of the view report contains the log of all the new reports along with the view option of AI suggestions and users' suggestions on the inappropriate and unacceptable language used in the content. Upon clicking one report, the AI-LT admin can watch detailed suggestions by users and AI on separate pages. They can accept, reject, and edit the suggestions before commanding AI to learn about the language for future suggestions. By clicking on the menu, the AI-LT can also view profiles of new AI-LT applicants and vote for them. 

To functionalize G4, \textit{Socheton} offers the users: (ci) play a quiz/game, (cii) reputation map, (ciii) personal records, (civ) profile set up. Playing quizzes consists of a three-page process where the user can challenge content or explanations on the second and third pages and report them for feedback. Users can choose if this report is about misinformation, inappropriate language, or both. By clicking the menu, they can view their personal record and track their progress of being conscious. The app also allows users to set up their profiles.

\textit{(D) Server. }
The server holds and processes the data in the application. The application sends the reports and feedback by the users and admins to the server. The server sorts and stores the specifically commanded ones by admins in its information bank for AI's use.

%The server posts new content and quiz-based games with answers and explanations prepared by AI and approved by admins. 

%\subsection{Implementation}

\begin{comment}

\begin{figure*}[!t]
\centering
\includegraphics[width=0.95\textwidth]{c5}
    \caption{Work-flow diagram of \textit{Unmochon} application with three major components: plugin at the user-end, the server with end-to-end encryption, and the customized Facebook group.}
    \label{fig:figure1}
\end{figure*}

\begin{figure*}[!t]
\centering
\includegraphics[width=0.95\textwidth]{c4}
    \caption{\textit{Unmochon} application takes a screenshot of Chrome tab that is currently open. It also copies the harasser's unique Facebook id from the address-bar. Before sending it to the database and posting to the dedicated Facebook page for \textit{Unmochon}, the application window only allows the users to mark the photo with red ink. (All the purple geometric shapes in the figure were post-edited for anonymization.)}
    \label{fig:figure1}
\end{figure*}

\end{comment}

%\newpage
\section{Phase-2: Users' Feedback on \textit{Socheton}}

\begin{figure*}[!t]
\centering
    \includegraphics[width=0.99\textwidth]{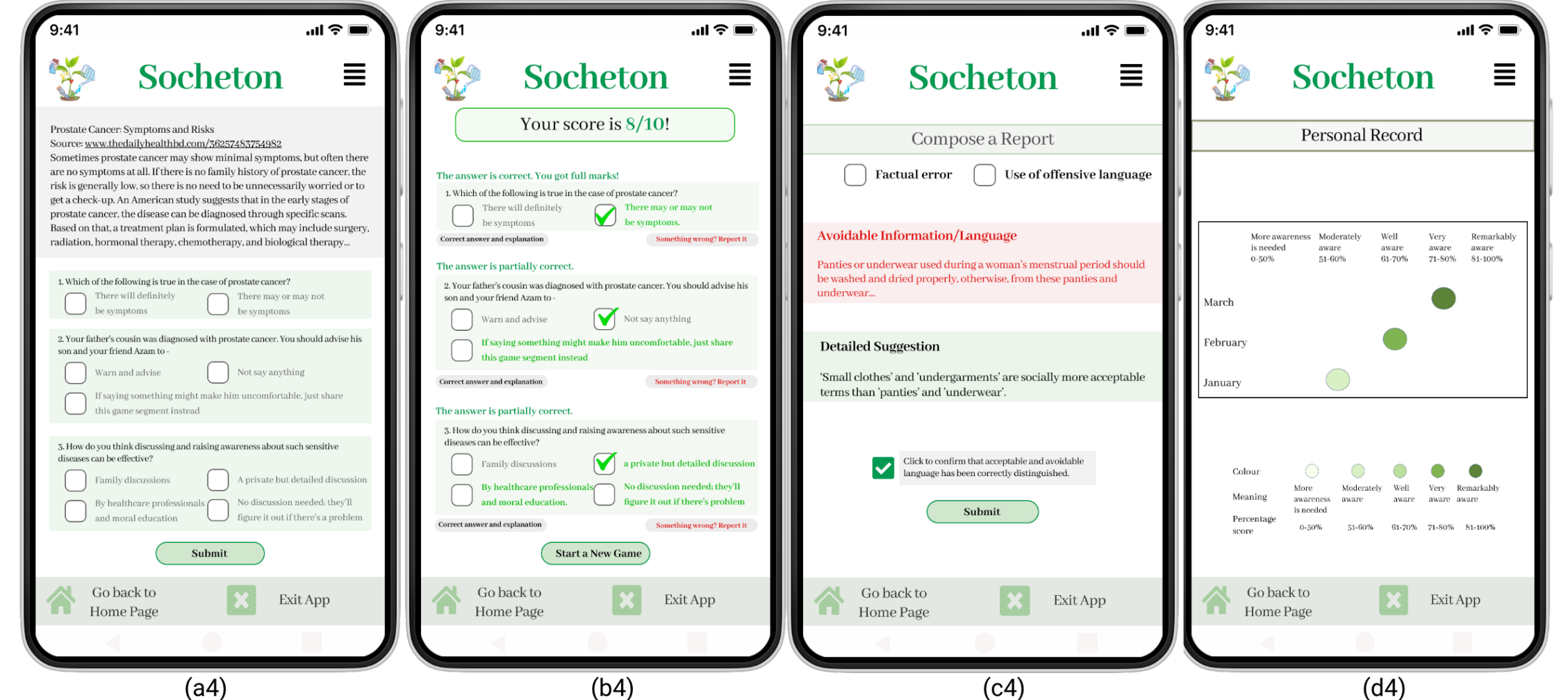}
    \caption{Example pages from \textit{Socheton} Application's User End. (a4) The quiz page, (b4) the score page that also lets them create a report, (c4) page where user can compose reports and add suggestions on information and language error cases, (d4) users' personal progress record.}
    \label{fig:figure1}
\end{figure*}

Our findings from Phase 1 suggest that reproductive well-being is a sensitive issue; people often shy away and avoid discussions in public. So, we were wary of conducting an intervention with the participants, rather we decided to share our prototype with the users with no active contact with the server and the human professionals and AI-teachers, explain the whole idea, and seek their feedback. We first prepared a user-study package, which included a prototype version of the `Socheton' and the user guide to the application. To avoid unintended stigma and withdrawal during the user study, \newedits{we shared the Bengali version of the prototype with no connection to the server and professionals and AI-teachers. Therefore, the components had no live data-sharing at the time of the user study.} For their better understanding, we also created a video demonstrating all the design goals and explaining how they will be functional in the application, and we added that video to the package that we shared with the participants. This phase of study design followed Mahar et al. \cite{mahar2018squadbox}. This part of the project was conducted during January-August 2024. Some participation happened in person, and some of them were online, based on the participant's convenience.

\subsection{Methods Used to Collect User-Feedback}
\newedits{In phase-2, we conducted focus group discussions and interviews with participants who lived in, originated from, or are currently living in Jessore.} We discuss the details of our methods for this phase below:

\subsubsection{Focus Group Discussion (FGD)}
We conducted four FGD sessions with 23 (18 female and five male) participants. Among them, thirteen previously participated in phase-1, and others were recruited with the help of RRF in a similar way as phase-1. Two participants were in the medical profession, and one was experienced in AI and language research. Upon reaching out to them, we explained the purpose of the work. We grouped relatively known participants of the same gender together and set up FGDs with them. We also sent them the user-study package a few hours before the sessions so that they could experience it and get some time to think about it. In the FGDs, we explained the application's purpose and helped them go through it step-by-step if needed. Then we discussed the benefits this application might bring to support such a stigmatized topic as reproductive well-being. We also asked them what challenges and troubles this application might generate for the users and other stakeholders.

The sessions were conducted in the Bengali language. The average length of the sessions was 35 minutes. Five participants gave us permission to audio record the sessions. We also offered them to discard the discussion at any moment due to their discomfort. However, no such incident happened. Participation in FGD was voluntary, and the participants were not paid.

\subsubsection{Interviews}
Along with the FGDs, we conducted eight interviews (five females and three males) with participants as they requested a one-on-one conversation instead of a group discussion. None of them were from the set of participants in pre-design interviews. As before, we recruited the rest of the participants from our professional networks on social media. Three of them were in the medical profession, and two were experienced in AI and language research. Upon reaching out to them, we explained the purpose of the work. We also sent them the user-study package a few hours before the interview so that they could experience it and get some time to think about it. During the interviews, we explained to them the application's purpose and helped them go through it step-by-step if needed. Then we discussed the benefits this application might bring to support such a stigmatized topic as reproductive well-being. We also asked them what challenges and troubles this application might generate for the users and other stakeholders.

All of the interview sessions used Bengali as the primary language. It generally took around 25 and 40 minutes to complete the interview. Three interviewees allowed us to audio-record their sessions. We left every opportunity for the participants to leave the interview if they felt uncomfortable, even during ongoing sessions. We also informed them that we would discard the record of their participation if they wanted. However, no such event took place throughout the study. Participation in the interview was voluntary, and the participants were not paid.

\begin{table}[!t]
%\begin{wraptable}{r}
\centering
%\begin{adjustbox}{max width=0.5\textwidth}
\begin{tabular}{|rl|}
\hline
Total Participants: & 28 (Female: 20, Male: 8)\\
\hdashline
\multicolumn{2}{|c|}{\textbf{Type of Participation}} \\
\hdashline
Interview only: & 5 (Female: 2, Male: 3)\\
FGD only: & 20 (Female: 15, Male: 5)\\
Interview + FGD: & 3 (Female: 3, Male: 0)\\
\hdashline

\multicolumn{2}{|c|}{\textbf{Age range (in Years)}} \\
\hdashline
All: & 19-60, median 33\\
Female: & 19-60, median 30\\
Male: & 19-56, median 38\\
\hdashline

\multicolumn{2}{|c|}{\textbf{Education}} \\
\hdashline
No Formal Schooling: & 5 (Female: 5, Male: 0)\\
Primary School: & 5 (Female: 5, Male: 0)\\
Secondary School: & 5 (Female: 2, Male: 3)\\
Higher Secondary School: & 4 (Female: 2, Male: 2)\\
Undergraduate and Above: & 9 (Female: 6, Male: 3)\\
\hline
\end{tabular}
%\end{adjustbox}
\caption{Demographics of User Study Participants}
%\vspace{-25pt}
%\end{wraptable}
\end{table}

\subsection{Data Collection and Analysis}
The audio files of interviews were recorded on the researchers' phones and later saved in a hard drive for further data processing steps. We collected approximately five hours of audio recordings and around 60 pages of field notes. The audio recordings were transcribed and translated into English. We then performed thematic analysis on the transcriptions and our detailed notes \cite{75strauss1990open, 76boyatzis1998transforming}. The authors independently read through the transcripts carefully and allowed codes to develop. Later they shared their codes with each other. A total of 18 codes spontaneously developed during the first round of coding. Then we clustered related codes into themes after a few iterations. Some of the themes seemed recurring, for example, religion, privacy, democracy, posting, household leadership, justice, etc. Such themes influence the organization of our findings section.

%\newpage
\section{Findings from Users' Feedback}
\textit{Socheton} met all the design goals as the prototype aimed at combating misinformation and misconception (G1), with G2 and G3 (culturally appropriate language and accountable moderation), and creating opportunities for users' democratic participation. Our users' feedback, concerns, and suggestions are discussed below.

\subsection{Design Concerns and Suggestions}
Ten participants anticipated that once such a tool is available, the server will be flooded with reports and many of those might be just random. They also proposed that this tool needs an intelligent filter to select and let in topic-related posts only, and flag the users that abuse the system. Also, the thirteen participants discussed the need for a guideline to define reproductive well-being concepts and logs for locally acceptable languages. They suggested running a nationwide survey asking the locals to help collect the acceptable language.

One design concern during an FGD session with eight participants was whether people would be motivated to use this application if launched in mass level, as they emphasized that this topic is already stigmatized and people generally want to shy away from it, as one of them explained, 

\begin{quote}
\textit{``Unless they suffer themselves, they would not even take the family members' (reproductive well-being related) problems seriously and ask them about it. They all will pose elegantly and call this a nasty topic to avoid discussions. Maybe more anonymous experience sharing here with the name of the area can help people grow interest and confidence to start thinking that thinking about these problems is important." (P78, Female, FGD-12)}
\end{quote}

This group further discussed that such experience-sharing with the location being named will induce concerns among people that they will know other people in their areas are also troubled, so such conversations should happen. Six participants suggested involving the local mosques as a stakeholder in the design. 

While discussing the part about authenticity, the participants said it was wise to add the doctors to the moderator panel to verify the information and have them give feedback on people's concerns with reference to sources. However, this discussion found new challenges with access, interpretability, and trustworthiness,

\begin{quote}
\textit{``If the resources they add in English and require us to purchase a subscription of 10 dollars or something, how can we do that? Neither do I understand technical or medical terms in English, nor will I buy. I will just ignore it. Also, I would like to know which doctor verified the information. Some doctors in the neighborhood are not good enough; we often see fraud doctors on the news that forged certificates to practice." (P70, Female, FGD-11)}
\end{quote}

\subsection{Reproductive Justice Perceptions and Concerns}
While our design did not solve most of the problems associated with reproductive well-being in the community, it raised some pivotal questions and challenged the existing practices of participation and empowerment in the reproductive well-being ecosystem.

\subsubsection{Combating Patriarchal Practices}
While our design intended to involve the community, eleven participants opined that there is still room for improvement in that area. For example, one FGD discussed that generally, conjugal lives are dictated by men, and household decisions are taken by in-laws; existing patriarchal and misogynic practices will challenge this and any other ICT-based interventions if they emphasize individual learning. One of them said,

\begin{quote}
\textit{``Ok, if I learn, I learn. But my bedroom is dictated by my husband, and the household by my father-in-law. How will the practices change if they do not learn? And how will they learn, as they are not ready to listen to me, a woman? And who is going to tell them that they should also learn from this app? It is also a matter of male ego." (P77, Female, FGD-12)}
\end{quote}

In a similar conversation with a doctor participant, they pointed out that the introduction to protection materials initiatives in many areas in Bangladesh benefited by involving sex workers, as they taught the men first, and thus, restricting AIDS and population control agendas also benefited. 

\subsubsection{Playing Democratic Participation}
Nine participants were concerned about system flooding and how that will play the democratic participation agenda of this tool. Five participants referenced Facebook's algorithm and explained how \textit{Socheton} might also fail. He brought our attention to advancing AI's and HCI-design's by employing more sophisticated democratic strategies in algorithmic practices to protect marginalized voices, 

\begin{quote}
\textit{``If this is open to all, people will abuse the reporting. Some will report western-medical-knowledge-based things, and some will report faith- and religion-inclusive contents. Depending on the area, the timeline of the use-end might be western-medical-knowledge heavy and faith- and religion-based content heavy, like the Facebook algorithm does to people's timelines. Is there any way that your AI can protect the actual sentiment of democratic participation by addressing actual minority voices?" (P91, Male, Interview)}
\end{quote}

Three participants had experience working with algorithms and natural languages, and they suggested learning about the Bangladeshi practice of sabotaging algorithms so that better strategies for protecting the platform's democratic participation agenda can be designed.  

\subsubsection{Weaponizing Identity Politics}
Seven participants mentioned that the problems are not only about technology or medical knowledge but also about the political complexities of the space. One participant who was a doctor referred to the recent event where people burnt high school book chapters with reproductive well-being content for having LGTBQ concepts in there, and explained how it weaponized many other different social issues, she said,

\begin{quote}
\textit{``One side of the problem is that they believed "concepts of LGBTQ should not be mentioned in high school books. The bigger part of the problem is they believed they could terrorize and patronize the nation's reproductive awareness by Islamizing --- like using religion as a weapon to downplay actual problems--- and politicizing the space by using their identity as Muslim males, foreign degree holders, intellectuals, etc. Well, in that case, no app can change if actual grassroots social movement and policy lever revisions do not take place, like the way the government has enforced banishment and punishment for gender reveal of the child in the mother's womb using hospital ultrasound facilities." (P68, Female, Interview)}
\end{quote}

Several other participants' opinions echoed hers, and they were dismissive of the possibilities that any intervention can alone make. They emphasized that for a broader impact, there should be more accountable participation of stakeholders in ensuring high-quality reproductive well-being for this highly sensitive community. 

\subsection{Feedback on User Interface and Process}
The participants gave us feedback on the User Interface of \textit{`Socheton'} during the user evaluation. For example, eight FGD participants said there should be sort options among the suggested quiz games based on the dates they were added and how many people have been playing to indicate which of them are popular. Eleven participants suggested converting it into a real-world mobile application and deploying it nationwide with ads being added to popular social and business platforms. Additionally, nine participants said the quiz-based games should include points for incentives, which could be cashed in e-market or e-business so that people are driven to use it. Ten participants insisted on making the application more informative, interactive, and guided. A participant explained her problem, 

\begin{quote}
\textit{``Wait, I am lost. Did the game end yet? You know what, there should have been a bar on top or below to indicate how much progress was made with the game because I am not sure how far I am now and how much more to go." (P62, Female, Interview)}
\end{quote}

Our prototype was created on Figma \cite{Figma:online}. During the feedback interview and FGDs, the Figma page did not load on the Google Chrome Browser on the researcher's phone. They had to switch to another browser or a computer to continue the study. However, this problem arose because of a mismatch between the phone's Android version, Chrome browser's version, and Figma's required GPU configuration. A total of 14 participants faced this issue during their sessions and suggested that our design should cope with such challenges while designing and deploying a real system.  
%\newpage
\section{Discussion}
This two-phase research investigated different aspects of reproductive well-being in Bangladeshi communities sensitive to faith- and value-based systems. In phase-1, our ethnography investigated the impact of sociocultural landscape, cultural beliefs, and healthcare infrastructure on Bangladeshi people's access to reproductive healthcare and set four design goals: combating misinformation, including culturally appropriate language, professionals accountable moderation, and promoting users' democratic participation. Building on `\textit{Distributive Justice,}' we designed  \textit{`Socheton,'} a culturally appropriate AI-mediated tool for reproductive well-being that includes healthcare professionals, AI-language teachers, and community members to moderate and run the activity-based tool and conducted user evaluation (n=28) in phase-2. Our findings from both phases open discussion on both design implication and theory ends. 

\subsection{Design Implications}
Our research generates several design implications for well-being-HCI and culturally appropriate AI design. We found opportunities for HCI-design's existing emphasis on health and well-being with traditional spiritual practices. Our participants pointed to religious institutions' existing role in reproductive well-being education and called for their inclusion as stakeholders. This calls for adopting participatory strategies involving reproductive well-being practitioners and HCI and AI designers collaborating with faith-based institutes and practitioners to support the community's reproductive well-being. A major finding of this research was language discrepancies that lead to resistance against modern concepts and methods in reproductive care. Participants suggested involving local community members to help HCI and AI understand language sensitivities in this space. Following the participant's suggestions, we can conduct a nationwide survey to accumulate acceptable language to help design culturally appropriate AI for reproductive well-being. Such design implications can join HCI's existing literature that combats local resistance against modern medicine and solicits situated and low-cost medicines and therapies for complex cultural and resource-constrained social settings \cite{94singh2010exploring, 95rastogi2010building, bidwell2015preserving}. Furthermore, designing theories and practices of culturally appropriate well-being-AI will benefit the ongoing sustainable-HCI movement by showing various ways to integrate design with local history, culture, environments, and politics \cite{97disalvo2010mapping, 98silberman2014next}. Building on Bidwell's work on integrating Indigenous knowledge with HCI design, we argue that rural Bangladeshi reproductive well-being should be scrutinized using the lens of local culture and its interpretations \cite{bidwell2015preserving}. Such a bridge between modern technologies and traditional cultural knowledge is a dire need in today's HCI4D scholarship \cite{99dearden2012see, 100dearden2015ethical, 101dodson2012ethics}.

\subsection{Broader Implications}
Our research also brings insight into some broader implications, which we discuss below. We also highly recommend addressing them while designing culturally appropriate AI-mediated reproductive well-being supporting tools, technologies, and activism for this and similar communities that are sensitive to complex faith-based values. 

\subsubsection{Toward Reproductive Well-being for All Gender}
Our research joins the literature on gender in HCI4D and well-being HCI \cite{keyes2020reimagining, sultana2020parareligious, spiel2019patching, shen2023human, homewood2021tracing}. While most of today's reproductive well-being revolves around women's menstruation, pregnancy, and maternal care, as well as adolescent and menopause awareness and care \cite{critchley2020menstruation, hennegan2019women, tutia2019hci, backonja2021there}, they sideline the fact that the reproductive well-being of men and people with other gender identities in the community determines the community's and its women's reproductive well-being, too. Our research addresses this gap in the literature and shows the complexities in this space. Note that designing reproductive well-being technologies for men and women in a rural Bangladeshi society that is currently Muslim-dominated and has a history of remaining colonized by multiple cultures for several hundred years is extremely challenging as our findings showed that the value sensitivities are highly heterogeneous here and influenced by heteronormalizing discourses. Such faith-based sentiments prevent men from overcoming stigma and seeking proper knowledge and care. This situation is unique and unknown to outsiders if they do not understand the norms in Bangladeshi communities sensitive to values and faith-based practices. Most existing reproductive well-being in HCI looks into women's problems as individuals and often fails to understand that reproductive well-being is highly community-centric. \newedits{We bring attention to this matter and argue for \textit{De-stereotyping gender sensitivities} of reproductive well-being and pushing the domain's focus from "women-only" to all.} This process can only happen through more engagement with the communities through activities, design sessions, ethnography research, and the development and deployment of new theories and frameworks to address their stigma and spiritual and religious sentiments in reproductive well-being support designed for them. 

%A group of ICTD scholars has also shown that developing and deploying tools, technologies, and frameworks for reproductive well-being is particularly complicated in the Global South due to resource constraints, value sensitivities, and cultural complexities \cite{kumar2020taking}. 

%Additionally, our work has particularly noted that such stigma and hurdles also influence men and create difficulties for them in seeking appropriate care for their reproductive well-being. We found that men also shy away from seeking care when their wives want to see female doctors to seek help for their conjugal reproductive decision-making. This kind of situation is unique and unknown to outsiders of the community if they do not understand the norms in rural Bangladesh. While most of today's reproductive well-being in HCI and ICTD looks into women's problems as individuals, this literature frequently fails to understand that reproductive well-being is highly community-centric. Our research brings well-being HCI's and ICTD's attention to this matter and argues for un-gendering the sensitivities of reproductive well-being from "women-only" to all. 

\subsubsection{Need for Community Sensitive Content Moderation}
Our research contributes to the ongoing discourse on content moderation within HCI4D, social media studies, and social justice in computing. While existing literature frequently criticizes contemporary content moderation techniques' bias and cultural insensitivity \cite{ohlund2023social, sultana2022toleration, lyu2024got, scott2023trauma, kumar2021braving}, our findings corroborate these claims. Our findings noted that social media content moderation often fails to recognize the nuanced ways individuals navigate stigma and cultural taboos through their online communication. For instance, using euphemisms or local jargon to discuss sensitive topics was frequently misinterpreted and flagged as inappropriate, as our participants reported. We argue that addressing these shortcomings requires a more inclusive approach to content moderation, including the active involvement of local moderators in reviewing flagged posts and comments to foster a more democratic and culturally sensitive online environment.

\subsubsection{Western Scientific Values vs. Local Sentiment}
Our research aligns with the postcolonial critique within HCI that posits technology as a tool of continued colonial domination in the Global South. \cite{sultana2019witchcraft,adamu2023no,lazem2021challenges,birhane2020algorithmic, mohamed2020decolonial, bon2022decolonizing, kamran2023decolonizing, adams2021can}. Furthermore, contemporary research on reproductive well-being underscores the Western-centric nature of many ICT tools and medical procedures that may not align with local religious beliefs, social customs, and cultural traditions \cite{klausen2019reproductive, till2022community}. Our research found supporting evidence for this. For example, to benefit from many modern productive care tools, participants were required to violate their religious sentiments and local norms. Instead, the participants turned to local traditional knowledge and practices that understood their sensitivities and were adaptive to their norms, even though such methods were not as effective as modern methods. We argue that HCI has an ethical responsibility to cohere with traditional values, knowledge, language, and material practices to develop a valid and respectable parallel and/or alternative support for the reproductive well-being of these and other similar communities for decolonizing HCI research and ensuring historical justice.

\subsubsection{De-centering Design Focus, Social Reform, and Distributive Justice for Reproductive Well-being}
Our work also joins the HCI literature on social reformation and justice \cite{sultana2021unmochon, sultana2022shishushurokkha, fox2016exploring, corbett2019engaging, bates2018future, fox2017social, asad2019academic}. Traditional \textit{User-Centered Design} design process's major concentration on usability goals, user characteristics, environment, tasks, and workflow of a process has often been criticized as it might lose moral values while satisfying the users~\cite{keinonen2008user, edwards2003challenges, norman2005human, rifat2020religion, abras2004user, mao2005state, holtzblatt2004rapid}. This led many of today's designers to draw the design goals on societal values over manipulative capitalist models and/or neoliberal dreams~\cite{friedman2013value, friedman2002value, friedman1996value, nouwen2015value, davis2015value}. Our findings also clearly pointed out communal responsibilities' crucial role in broadly ensuring reproductive well-being and urged to de-center the design focus to the whole society by employing the model of \textit{Distributive Justice} in the design space \cite{Rescher66dis, sen1982utilitarianism, schmidtz1998social, schroth2008distributive}. 

Our read from the findings is that the fight against the stigma of reproductive well-being in Bangladesh will benefit more by not fighting against the socially prominent faith- and value-based sensitivities but instead making peace with them. While it would require a major social reformation for Western value-based reproductive well-being support systems to find their way to be effective in this space, that will not happen overnight. Our participants helped us find HCI-design's possible future directions by questioning our intent --- "Who is going to teach the others in the household?". We infer that reproductive well-being in Bangladesh needs the attention of design researchers who would invest in designing technologies to allow all the community members to contribute to the social reformation. Such a reformation would include growing awareness, organizing social support, and resisting the abuse of cultural values against well-being agendas. However, participation of all the community members might raise some derivative concerns, including resource constraints, unfair amount of extra workload, stress, etc. This calls for social reformation strategies to build on \textit{Distributive Justice} approach. Welfare-based principles of \textit{Distributive Justice} suggest that material goods and services in society should be distributed in a way that contributes to social welfare and further allows the members of the community to share the welfare responsibilities equitably \cite{schmidtz1998social, schroth2008distributive, sen1982utilitarianism, Rescher66dis}. We call for more strategic HCI design orientations to facilitate the abovementioned comprehensive societal transformation aligning with \textit{Distributive Justice} principles.

\subsubsection{Expanding Democracy within HCI4D and AI for Social Good}
Our work also joins the democracy in HCI scholarship. Most of today's HCI literature sees democratic participation in decision-making as an apparatus for preserving human rights, social harmony, and sustainability \cite{kelty2015seven, sultana2021unmochon, sultana2019witchcraft, le2012participation, asad2017tap}. This chain of work has discussed users' and community's participation in peer production, being part of technology-mediated collaboration, and being involved in event discussions via social media by providing opinions and feedback. Only a handful of research has looked into democratic participation in narrative making \cite{claes2017impact, sultana2020viz, sultana2023communicating}. However, most literature is still missing out on how to allow the communities in question to be active makers or creators and participate in driving the ecosystem. In our design, we followed the traditional HCI co-design style, discussed the design goals with the participants, sought their feedback, and prototyped \textit{Socheton}, which allowed them to provide feedback on reproductive well-being contents. However, the participants suggested letting them be the content creators alongside being feedback providers so that they can decide on culturally appropriate language, phasing, and values with more authenticity and accountability. They also suggested collecting locally acceptable language contents through surveys to develop culturally appropriate language tools to support this faith-sensitive community's reproductive well-being. Among a rare collection that aligns with this sentiment, a recent cultural visualization work by Sultana et al. adopted this approach, letting the participants create the content, and the community constructs the narrative \cite{sultana2023communicating}. This strategy dismisses the sentiment held by many scholars that marginalized communities may not know about well-being and quality of life, and hence, they would require Western knowledge-based guidelines, regardless, that suppress marginalized values, faith, language, and knowledge. Based on our findings, we position against this sentiment and argue for expanding the concept and practice of democracy in HCI4D and AI for social good to protect marginalized faiths, cultures, and knowledge.

\newedits{This urges further attention to power imbalances, accountability, and trustworthiness of information infrastructure in user-driven decision-making. In \textit{Socheton} prototype, power imbalances and misinformation in user-driven decision-making are planned to be managed through a multi-layered moderation system involving healthcare professionals, culturally aware AI-language teachers, and community-elected administrators. In actual deployment, this structure is expected to ensure accountable content evaluation, validation, and culturally sensitive moderation, while still empowering users to report and challenge content. The platform is designed to support democratic participation by enabling users to voice concerns and suggest changes, while mitigating misuse with AI filters and expert oversight. Participants emphasized the need for trustworthy verification, accessible language, and safeguards against algorithmic bias to protect marginalized voices and prevent dominant groups from monopolizing discourse. Future iterations of the system could incorporate personalized content filtering based on users' cultural and linguistic preferences to enhance inclusivity and trust. Additionally, integrating community-driven language banks and partnerships with local institutions like mosques, temples, and schools may further ground the tool in everyday social contexts and improve adoption. Thus, Socheton contributes to HCI, political design, and AI for social good theories by advancing how distributive justice frameworks can operationalize culturally sensitive, human-centered, and community-moderated AI systems that balance participatory agency with accountability in stigmatized domains.}

\subsection{Limitation and Future Work}
Our work has several limitations. First, since reproductive well-being is a stigmatized topic in Bangladesh, we recruited the participants through convenience sampling, and thus our exploratory work is not free from bias. However, this bias is important since it informs us about the existence, problematic nature, and damaging impact of ignoring reproductive well-being concerns in Bangladesh and hence, we do not make any generalized claim. Therefore, this approach aligned with HCI's sensibility of design from the margin by engaging with groups around and beyond the development and helped us understand the intersectional nature of the community  \cite{jain2021margins}. Another limitation of our work is that we did not engage in a deeper discussion with LGBTQ+ communities on this topic, although some survey responses pointed to this aspect of gender, as LGBTQ+ concepts are also highly taboo in the country, as we briefly explained in the findings. However, bringing in this important aspect would require more specific research questions and a carefully designed study, which was beyond the scope of our work. Despite these limitations, our study will be useful for designing technologies to support reproductive well-being in the context of communities highly sensitive to faith- and value-based systems in low-resource and patriarchal settings. Also, the arguments and lessons from this study will contribute to reproductive well-being policy-making in Bangladesh and other countries with similar resources, environments, and social settings.

In future work, we will include a nationwide survey to accumulate and develop a rich database of acceptable language in the community and release it for open access for local and global researchers to help design culturally appropriate tools and technologies to support the reproductive well-being of marginalized Bangladeshi communities. We are considering employing purposeful sampling to engage with specific target groups, e.g., low-income, LGBTQ+, people with disabilities, or other minority groups that have been overlooked before by research on reproductive justice. In addition, we plan to involve local AI and language researchers, medical scientists, community spirituality, and religious leaders who are influential in this space and collaborate in designing to combat the community's stigma of reproductive well-being. In this regard, we will seek collaboration with the local government, the national health ministry, different local and international NGOs, and policymakers.

\section{Conclusion}
This paper draws on a two-phase design research in Bangladesh that investigated different aspects of reproductive well-being with communities sensitive to faith- and value-based systems. The phase-1 ethnography researched the impact of sociocultural landscape, cultural beliefs, and healthcare infrastructure on Bangladeshi people's access to quality reproductive healthcare and set four design goals: combating misinformation, including culturally appropriate language, professionals accountable moderation, and promoting users' democratic participation. Drawing on the findings and `\textit{Distributive Justice,}' we designed \textit{`Socheton,'} a culturally appropriate reproductive well-being tool for collaborations among healthcare professionals, AI-language teachers, and community members, and conducted user evaluation (n=28) in phase-2. Our findings initiate discussions on the ethical and practical challenges surrounding designing reproductive well-being technologies for Bangladeshi communities sensitive to cultural and faith-based values and similar other communities marginalized in well-being HCI and HCI design scholarship.

\begin{acks}
\newedits{We thank the participants for generously sharing their time, experiences, and insights throughout the study. We also thank the Rural Reconstruction Foundation (RRF) for their crucial assistance with community access and participant recruitment. This work was supported by the Dicovery Partners Institute (DPI) of the University Illinois System and funded by the Sharifa Sultana's and Nervo Verdezoto's joint grant from Cardiff University-University of Illinois System Joint Research and Innovation Seed Grants Program, under grant number \#C1-F100025-O434040-P434558 and Sharifa Sultana's ICR fund \#C1-F200250-O434053-P434533.} 
\end{acks}
%\input{7.lim-fw-con}

%\newpage

%\newpage

\bibliographystyle{ACM-Reference-Format}
\bibliography{sample-base}

%%% -*-BibTeX-*-
%%% Do NOT edit. File created by BibTeX with style
%%% ACM-Reference-Format-Journals [18-Jan-2012].

\begin{thebibliography}{150}

%%% ====================================================================
%%% NOTE TO THE USER: you can override these defaults by providing
%%% customized versions of any of these macros before the \bibliography
%%% command.  Each of them MUST provide its own final punctuation,
%%% except for \shownote{}, \showDOI{}, and \showURL{}.  The latter two
%%% do not use final punctuation, in order to avoid confusing it with
%%% the Web address.
%%%
%%% To suppress output of a particular field, define its macro to expand
%%% to an empty string, or better, \unskip, like this:
%%%
%%% \newcommand{\showDOI}[1]{\unskip}   % LaTeX syntax
%%%
%%% \def \showDOI #1{\unskip}           % plain TeX syntax
%%%
%%% ====================================================================

\ifx \showCODEN    \undefined \def \showCODEN     #1{\unskip}     \fi
\ifx \showDOI      \undefined \def \showDOI       #1{#1}\fi
\ifx \showISBNx    \undefined \def \showISBNx     #1{\unskip}     \fi
\ifx \showISBNxiii \undefined \def \showISBNxiii  #1{\unskip}     \fi
\ifx \showISSN     \undefined \def \showISSN      #1{\unskip}     \fi
\ifx \showLCCN     \undefined \def \showLCCN      #1{\unskip}     \fi
\ifx \shownote     \undefined \def \shownote      #1{#1}          \fi
\ifx \showarticletitle \undefined \def \showarticletitle #1{#1}   \fi
\ifx \showURL      \undefined \def \showURL       {\relax}        \fi
% The following commands are used for tagged output and should be
% invisible to TeX
\providecommand\bibfield[2]{#2}
\providecommand\bibinfo[2]{#2}
\providecommand\natexlab[1]{#1}
\providecommand\showeprint[2][]{arXiv:#2}

\bibitem[\protect\citeauthoryear{Abras, Maloney-Krichmar, Preece, et~al\mbox{.}}{Abras et~al\mbox{.}}{2004}]%
        {abras2004user}
\bibfield{author}{\bibinfo{person}{Chadia Abras}, \bibinfo{person}{Diane Maloney-Krichmar}, \bibinfo{person}{Jenny Preece}, {et~al\mbox{.}}} \bibinfo{year}{2004}\natexlab{}.
\newblock \showarticletitle{User-centered design}.
\newblock \bibinfo{journal}{\emph{Bainbridge, W. Encyclopedia of Human-Computer Interaction. Thousand Oaks: Sage Publications}} \bibinfo{volume}{37}, \bibinfo{number}{4} (\bibinfo{year}{2004}), \bibinfo{pages}{445--456}.
\newblock


\bibitem[\protect\citeauthoryear{Adams}{Adams}{2021}]%
        {adams2021can}
\bibfield{author}{\bibinfo{person}{Rachel Adams}.} \bibinfo{year}{2021}\natexlab{}.
\newblock \showarticletitle{Can artificial intelligence be decolonized?}
\newblock \bibinfo{journal}{\emph{Interdisciplinary Science Reviews}} \bibinfo{volume}{46}, \bibinfo{number}{1-2} (\bibinfo{year}{2021}), \bibinfo{pages}{176--197}.
\newblock


\bibitem[\protect\citeauthoryear{Adamu}{Adamu}{2023}]%
        {adamu2023no}
\bibfield{author}{\bibinfo{person}{Muhammad~Sadi Adamu}.} \bibinfo{year}{2023}\natexlab{}.
\newblock \showarticletitle{No more “solutionism” or “saviourism” in futuring African HCI: A manyfesto}.
\newblock \bibinfo{journal}{\emph{ACM Transactions on Computer-Human Interaction}} \bibinfo{volume}{30}, \bibinfo{number}{2} (\bibinfo{year}{2023}), \bibinfo{pages}{1--42}.
\newblock


\bibitem[\protect\citeauthoryear{Akoth, Oguta, and Gatimu}{Akoth et~al\mbox{.}}{2021}]%
        {akoth2021prevalence}
\bibfield{author}{\bibinfo{person}{Catherine Akoth}, \bibinfo{person}{James~Odhiambo Oguta}, {and} \bibinfo{person}{Samwel~Maina Gatimu}.} \bibinfo{year}{2021}\natexlab{}.
\newblock \showarticletitle{Prevalence and factors associated with covert contraceptive use in Kenya: a cross-sectional study}.
\newblock \bibinfo{journal}{\emph{BMC public health}}  \bibinfo{volume}{21} (\bibinfo{year}{2021}), \bibinfo{pages}{1--8}.
\newblock


\bibitem[\protect\citeauthoryear{Al-Naimi and Alistar}{Al-Naimi and Alistar}{2024}]%
        {al2024understanding}
\bibfield{author}{\bibinfo{person}{Latifa Al-Naimi} {and} \bibinfo{person}{Mirela Alistar}.} \bibinfo{year}{2024}\natexlab{}.
\newblock \showarticletitle{Understanding Cultural and Religious Values Relating to Awareness of Women’s Intimate Health among Arab Muslims}. In \bibinfo{booktitle}{\emph{Proceedings of the CHI Conference on Human Factors in Computing Systems}}. \bibinfo{pages}{1--18}.
\newblock


\bibitem[\protect\citeauthoryear{Almeida}{Almeida}{2015}]%
        {almeida2015designing}
\bibfield{author}{\bibinfo{person}{Teresa Almeida}.} \bibinfo{year}{2015}\natexlab{}.
\newblock \showarticletitle{Designing intimate wearables to promote preventative health care practices}. In \bibinfo{booktitle}{\emph{Adjunct Proceedings of the 2015 ACM International Joint Conference on Pervasive and Ubiquitous Computing and Proceedings of the 2015 ACM International Symposium on Wearable Computers}}. \bibinfo{pages}{659--662}.
\newblock


\bibitem[\protect\citeauthoryear{Almeida, Balaam, and Comber}{Almeida et~al\mbox{.}}{2020}]%
        {almeida2020woman}
\bibfield{author}{\bibinfo{person}{Teresa Almeida}, \bibinfo{person}{Madeline Balaam}, {and} \bibinfo{person}{Rob Comber}.} \bibinfo{year}{2020}\natexlab{}.
\newblock \showarticletitle{Woman-centered design through humanity, activism, and inclusion}.
\newblock \bibinfo{journal}{\emph{ACM Transactions on Computer-Human Interaction (TOCHI)}} \bibinfo{volume}{27}, \bibinfo{number}{4} (\bibinfo{year}{2020}), \bibinfo{pages}{1--30}.
\newblock


\bibitem[\protect\citeauthoryear{Almeida, Comber, Wood, Saraf, and Balaam}{Almeida et~al\mbox{.}}{2016}]%
        {55almeida2016looking}
\bibfield{author}{\bibinfo{person}{Teresa Almeida}, \bibinfo{person}{Rob Comber}, \bibinfo{person}{Gavin Wood}, \bibinfo{person}{Dean Saraf}, {and} \bibinfo{person}{Madeline Balaam}.} \bibinfo{year}{2016}\natexlab{}.
\newblock \showarticletitle{On Looking at the Vagina through Labella}. In \bibinfo{booktitle}{\emph{Proceedings of the 2016 CHI Conference on Human Factors in Computing Systems}}. ACM, \bibinfo{pages}{1810--1821}.
\newblock


\bibitem[\protect\citeauthoryear{Aponjon}{Aponjon}{2019}]%
        {Aponjon14:online}
\bibfield{author}{\bibinfo{person}{Aponjon}.} \bibinfo{year}{2019}\natexlab{}.
\newblock \bibinfo{howpublished}{\url{http://www.aponjon.com.bd/}}.
\newblock


\bibitem[\protect\citeauthoryear{Arousell and Carlbom}{Arousell and Carlbom}{2016}]%
        {arousell2016culture}
\bibfield{author}{\bibinfo{person}{Jonna Arousell} {and} \bibinfo{person}{Aje Carlbom}.} \bibinfo{year}{2016}\natexlab{}.
\newblock \showarticletitle{Culture and religious beliefs in relation to reproductive health}.
\newblock \bibinfo{journal}{\emph{Best practice \& research Clinical obstetrics \& gynaecology}}  \bibinfo{volume}{32} (\bibinfo{year}{2016}), \bibinfo{pages}{77--87}.
\newblock


\bibitem[\protect\citeauthoryear{Asad, Dombrowski, Costanza-Chock, Erete, and Harrington}{Asad et~al\mbox{.}}{2019}]%
        {asad2019academic}
\bibfield{author}{\bibinfo{person}{Mariam Asad}, \bibinfo{person}{Lynn Dombrowski}, \bibinfo{person}{Sasha Costanza-Chock}, \bibinfo{person}{Sheena Erete}, {and} \bibinfo{person}{Christina Harrington}.} \bibinfo{year}{2019}\natexlab{}.
\newblock \showarticletitle{Academic accomplices: Practical strategies for research justice}. In \bibinfo{booktitle}{\emph{Companion Publication of the 2019 on Designing Interactive Systems Conference 2019 Companion}}. \bibinfo{pages}{353--356}.
\newblock


\bibitem[\protect\citeauthoryear{Asad and Le~Dantec}{Asad and Le~Dantec}{2017}]%
        {asad2017tap}
\bibfield{author}{\bibinfo{person}{Mariam Asad} {and} \bibinfo{person}{Christopher~A Le~Dantec}.} \bibinfo{year}{2017}\natexlab{}.
\newblock \showarticletitle{Tap the" make this public" button: A design-based inquiry into issue advocacy and digital civics}. In \bibinfo{booktitle}{\emph{Proceedings of the 2017 CHI Conference on Human Factors in Computing Systems}}. \bibinfo{pages}{6304--6316}.
\newblock


\bibitem[\protect\citeauthoryear{Ataullahjan, Mumtaz, and Vallianatos}{Ataullahjan et~al\mbox{.}}{2019}]%
        {ataullahjan2019family}
\bibfield{author}{\bibinfo{person}{Anushka Ataullahjan}, \bibinfo{person}{Zubia Mumtaz}, {and} \bibinfo{person}{Helen Vallianatos}.} \bibinfo{year}{2019}\natexlab{}.
\newblock \showarticletitle{Family planning in Pakistan: A site of resistance}.
\newblock \bibinfo{journal}{\emph{Social Science \& Medicine}}  \bibinfo{volume}{230} (\bibinfo{year}{2019}), \bibinfo{pages}{158--165}.
\newblock


\bibitem[\protect\citeauthoryear{Axelsen}{Axelsen}{1985}]%
        {axelsen1985women}
\bibfield{author}{\bibinfo{person}{Diana~E Axelsen}.} \bibinfo{year}{1985}\natexlab{}.
\newblock \showarticletitle{Women as victims of medical experimentation: J. Marion Sims' surgery on slave women, 1845-1850}.
\newblock \bibinfo{journal}{\emph{Sage}} \bibinfo{volume}{2}, \bibinfo{number}{2} (\bibinfo{year}{1985}), \bibinfo{pages}{10}.
\newblock


\bibitem[\protect\citeauthoryear{Backonja, Taylor-Swanson, Miller, Jung, Haldar, and Woods}{Backonja et~al\mbox{.}}{2021}]%
        {backonja2021there}
\bibfield{author}{\bibinfo{person}{Uba Backonja}, \bibinfo{person}{Lisa Taylor-Swanson}, \bibinfo{person}{Andrew~D Miller}, \bibinfo{person}{Se-Hee Jung}, \bibinfo{person}{Shefali Haldar}, {and} \bibinfo{person}{Nancy~Fugate Woods}.} \bibinfo{year}{2021}\natexlab{}.
\newblock \showarticletitle{“There’sa problem, now what’s the solution?”: suggestions for technologies to support the menopausal transition from individuals experiencing menopause and healthcare practitioners}.
\newblock \bibinfo{journal}{\emph{Journal of the American Medical Informatics Association}} \bibinfo{volume}{28}, \bibinfo{number}{2} (\bibinfo{year}{2021}), \bibinfo{pages}{209--221}.
\newblock


\bibitem[\protect\citeauthoryear{Bagalkot, Akbar, Sharma, Mackintosh, Harrington, Griffiths, Noronha, and Verdezoto}{Bagalkot et~al\mbox{.}}{2022}]%
        {bagalkot2022embodied}
\bibfield{author}{\bibinfo{person}{Naveen Bagalkot}, \bibinfo{person}{Syeda~Zainab Akbar}, \bibinfo{person}{Swati Sharma}, \bibinfo{person}{Nicola Mackintosh}, \bibinfo{person}{Deirdre Harrington}, \bibinfo{person}{Paula Griffiths}, \bibinfo{person}{Judith~Angelitta Noronha}, {and} \bibinfo{person}{Nervo Verdezoto}.} \bibinfo{year}{2022}\natexlab{}.
\newblock \showarticletitle{Embodied negotiations, practices and experiences interacting with pregnancy care infrastructures in South India}. In \bibinfo{booktitle}{\emph{Proceedings of the 2022 CHI Conference on Human Factors in Computing Systems}}. \bibinfo{pages}{1--21}.
\newblock


\bibitem[\protect\citeauthoryear{Bardzell, Bardzell, Lazar, and Su}{Bardzell et~al\mbox{.}}{2019}]%
        {bardzell2019re}
\bibfield{author}{\bibinfo{person}{Jeffrey Bardzell}, \bibinfo{person}{Shaowen Bardzell}, \bibinfo{person}{Amanda Lazar}, {and} \bibinfo{person}{Norman~Makoto Su}.} \bibinfo{year}{2019}\natexlab{}.
\newblock \showarticletitle{(Re-) Framing menopause experiences for HCI and design}. In \bibinfo{booktitle}{\emph{proceedings of the 2019 CHI Conference on Human Factors in Computing Systems}}. \bibinfo{pages}{1--13}.
\newblock


\bibitem[\protect\citeauthoryear{Bates, Thomas, Remy, Nathan, Mann, and Friday}{Bates et~al\mbox{.}}{2018}]%
        {bates2018future}
\bibfield{author}{\bibinfo{person}{Oliver Bates}, \bibinfo{person}{Vanessa Thomas}, \bibinfo{person}{Christian Remy}, \bibinfo{person}{Lisa~P Nathan}, \bibinfo{person}{Samuel Mann}, {and} \bibinfo{person}{Adrian Friday}.} \bibinfo{year}{2018}\natexlab{}.
\newblock \showarticletitle{The Future of HCI and Sustainability: Championing Environmental and Social Justice.}. In \bibinfo{booktitle}{\emph{Extended Abstracts of the 2018 CHI Conference on Human Factors in Computing Systems}}. \bibinfo{pages}{1--4}.
\newblock


\bibitem[\protect\citeauthoryear{Biernacki and Waldorf}{Biernacki and Waldorf}{1981}]%
        {Biernacki1981}
\bibfield{author}{\bibinfo{person}{P. Biernacki} {and} \bibinfo{person}{D. Waldorf}.} \bibinfo{year}{1981}\natexlab{}.
\newblock \showarticletitle{{Snowball Sampling: Problems and Techniques of Chain Referral Sampling}}.
\newblock \bibinfo{journal}{\emph{Sociological Methods {\&} Research}}  \bibinfo{volume}{10} (\bibinfo{year}{1981}), \bibinfo{pages}{141--163}.
\newblock


\bibitem[\protect\citeauthoryear{Birhane}{Birhane}{2020}]%
        {birhane2020algorithmic}
\bibfield{author}{\bibinfo{person}{Abeba Birhane}.} \bibinfo{year}{2020}\natexlab{}.
\newblock \showarticletitle{Algorithmic colonization of Africa}.
\newblock \bibinfo{journal}{\emph{SCRIPTed}}  \bibinfo{volume}{17} (\bibinfo{year}{2020}), \bibinfo{pages}{389}.
\newblock


\bibitem[\protect\citeauthoryear{Blackwell, Dimond, Schoenebeck, and Lampe}{Blackwell et~al\mbox{.}}{2017}]%
        {blackwell2017classification}
\bibfield{author}{\bibinfo{person}{Lindsay Blackwell}, \bibinfo{person}{Jill Dimond}, \bibinfo{person}{Sarita Schoenebeck}, {and} \bibinfo{person}{Cliff Lampe}.} \bibinfo{year}{2017}\natexlab{}.
\newblock \showarticletitle{Classification and its consequences for online harassment: Design insights from heartmob}.
\newblock \bibinfo{journal}{\emph{Proceedings of the ACM on Human-Computer Interaction}} \bibinfo{volume}{1}, \bibinfo{number}{CSCW} (\bibinfo{year}{2017}), \bibinfo{pages}{1--19}.
\newblock


\bibitem[\protect\citeauthoryear{Bon, Dittoh, L{\^o}, Pini, Bwana, Cheah, Kulathuramaiyer, and Baart}{Bon et~al\mbox{.}}{2022}]%
        {bon2022decolonizing}
\bibfield{author}{\bibinfo{person}{Anna Bon}, \bibinfo{person}{Francis Dittoh}, \bibinfo{person}{Gossa L{\^o}}, \bibinfo{person}{M{\'o}nica Pini}, \bibinfo{person}{Robert~Masua Bwana}, \bibinfo{person}{WaiShiang Cheah}, \bibinfo{person}{Narayanan Kulathuramaiyer}, {and} \bibinfo{person}{Andr{\'e} Baart}.} \bibinfo{year}{2022}\natexlab{}.
\newblock \bibinfo{title}{Decolonizing Technology and Society: A Perspective from the Global South.}
\newblock
\newblock


\bibitem[\protect\citeauthoryear{Boyatzis}{Boyatzis}{1998}]%
        {76boyatzis1998transforming}
\bibfield{author}{\bibinfo{person}{Richard~E Boyatzis}.} \bibinfo{year}{1998}\natexlab{}.
\newblock \bibinfo{booktitle}{\emph{Transforming qualitative information: Thematic analysis and code development}}.
\newblock \bibinfo{publisher}{sage}.
\newblock


\bibitem[\protect\citeauthoryear{Bracke}{Bracke}{2022}]%
        {bracke2022women}
\bibfield{author}{\bibinfo{person}{Maud~Anne Bracke}.} \bibinfo{year}{2022}\natexlab{}.
\newblock \showarticletitle{Women’s rights, family planning, and population control: the emergence of reproductive rights in the united nations (1960s--70s)}.
\newblock \bibinfo{journal}{\emph{The International History Review}} \bibinfo{volume}{44}, \bibinfo{number}{4} (\bibinfo{year}{2022}), \bibinfo{pages}{751--771}.
\newblock


\bibitem[\protect\citeauthoryear{Branding}{Branding}{2019a}]%
        {khulna:online}
\bibfield{author}{\bibinfo{person}{Zilla Branding}.} \bibinfo{year}{2019}\natexlab{a}.
\newblock \bibinfo{howpublished}{\url{http://www.khulna.gov.bd/}}.
\newblock


\bibitem[\protect\citeauthoryear{Branding}{Branding}{2019b}]%
        {Jessore1:online}
\bibfield{author}{\bibinfo{person}{Zilla Branding}.} \bibinfo{year}{2019}\natexlab{b}.
\newblock \bibinfo{howpublished}{\url{ http://www.jessore.gov.bd/}}.
\newblock


\bibitem[\protect\citeauthoryear{Britton, Tumlinson, Williams, Gorrindo, Onyango, and Wambua}{Britton et~al\mbox{.}}{2021}]%
        {britton2021women}
\bibfield{author}{\bibinfo{person}{Laura~E Britton}, \bibinfo{person}{Katherine Tumlinson}, \bibinfo{person}{Caitlin~R Williams}, \bibinfo{person}{Phillip Gorrindo}, \bibinfo{person}{Dickens Onyango}, {and} \bibinfo{person}{Debborah Wambua}.} \bibinfo{year}{2021}\natexlab{}.
\newblock \showarticletitle{How women and providers perceive male partner resistance to contraceptives in Western Kenya: a qualitative study}.
\newblock \bibinfo{journal}{\emph{Sexual \& Reproductive Healthcare}}  \bibinfo{volume}{29} (\bibinfo{year}{2021}), \bibinfo{pages}{100650}.
\newblock


\bibitem[\protect\citeauthoryear{Campo~Woytuk, Nilsson, and Liu}{Campo~Woytuk et~al\mbox{.}}{2019}]%
        {campo2019your}
\bibfield{author}{\bibinfo{person}{Nadia Campo~Woytuk}, \bibinfo{person}{Linette Nilsson}, {and} \bibinfo{person}{Mingxing Liu}.} \bibinfo{year}{2019}\natexlab{}.
\newblock \showarticletitle{Your period rules: Design implications for period-positive technologies}. In \bibinfo{booktitle}{\emph{Extended Abstracts of the 2019 CHI Conference on Human Factors in Computing Systems}}. \bibinfo{pages}{1--6}.
\newblock


\bibitem[\protect\citeauthoryear{Campo~Woytuk, S{\o}ndergaard, Ciolfi~Felice, and Balaam}{Campo~Woytuk et~al\mbox{.}}{2020}]%
        {campo2020touching}
\bibfield{author}{\bibinfo{person}{Nadia Campo~Woytuk}, \bibinfo{person}{Marie Louise~Juul S{\o}ndergaard}, \bibinfo{person}{Marianela Ciolfi~Felice}, {and} \bibinfo{person}{Madeline Balaam}.} \bibinfo{year}{2020}\natexlab{}.
\newblock \showarticletitle{Touching and being in touch with the menstruating body}. In \bibinfo{booktitle}{\emph{Proceedings of the 2020 CHI conference on human factors in computing systems}}. \bibinfo{pages}{1--14}.
\newblock


\bibitem[\protect\citeauthoryear{Casterline, Sathar, and ul~Haque}{Casterline et~al\mbox{.}}{2001}]%
        {casterline2001obstacles}
\bibfield{author}{\bibinfo{person}{John~B Casterline}, \bibinfo{person}{Zeba~A Sathar}, {and} \bibinfo{person}{Minhaj ul Haque}.} \bibinfo{year}{2001}\natexlab{}.
\newblock \showarticletitle{Obstacles to contraceptive use in Pakistan: A study in Punjab}.
\newblock \bibinfo{journal}{\emph{Studies in family planning}} \bibinfo{volume}{32}, \bibinfo{number}{2} (\bibinfo{year}{2001}), \bibinfo{pages}{95--110}.
\newblock


\bibitem[\protect\citeauthoryear{Chhachhi}{Chhachhi}{1989}]%
        {chhachhi1989state}
\bibfield{author}{\bibinfo{person}{Amrita Chhachhi}.} \bibinfo{year}{1989}\natexlab{}.
\newblock \showarticletitle{The state, religious fundamentalism and women: Trends in South Asia}.
\newblock \bibinfo{journal}{\emph{Economic and Political Weekly}} (\bibinfo{year}{1989}), \bibinfo{pages}{567--578}.
\newblock


\bibitem[\protect\citeauthoryear{Chordia, Baltaxe-Admony, Boone, Sheehan, Dombrowski, Le~Dantec, Ringland, and Smith}{Chordia et~al\mbox{.}}{2024}]%
        {chordia2024social}
\bibfield{author}{\bibinfo{person}{Ishita Chordia}, \bibinfo{person}{Leya~Breanna Baltaxe-Admony}, \bibinfo{person}{Ashley Boone}, \bibinfo{person}{Alyssa Sheehan}, \bibinfo{person}{Lynn Dombrowski}, \bibinfo{person}{Christopher~A Le~Dantec}, \bibinfo{person}{Kathryn~E Ringland}, {and} \bibinfo{person}{Angela~DR Smith}.} \bibinfo{year}{2024}\natexlab{}.
\newblock \showarticletitle{Social Justice in HCI: A Systematic Literature Review}. In \bibinfo{booktitle}{\emph{Proceedings of the CHI Conference on Human Factors in Computing Systems}}. \bibinfo{pages}{1--33}.
\newblock


\bibitem[\protect\citeauthoryear{Claes and Vande~Moere}{Claes and Vande~Moere}{2017}]%
        {claes2017impact}
\bibfield{author}{\bibinfo{person}{Sandy Claes} {and} \bibinfo{person}{Andrew Vande~Moere}.} \bibinfo{year}{2017}\natexlab{}.
\newblock \showarticletitle{The impact of a narrative design strategy for information visualization on a public display}. In \bibinfo{booktitle}{\emph{Proceedings of the 2017 Conference on Designing Interactive Systems}}. \bibinfo{pages}{833--838}.
\newblock


\bibitem[\protect\citeauthoryear{Coleman, Till, Farao, Shandu, Khuzwayo, Muthelo, Mbombi, Bopape, van Heerden, Mothiba, et~al\mbox{.}}{Coleman et~al\mbox{.}}{2023}]%
        {coleman2023reconsidering}
\bibfield{author}{\bibinfo{person}{Toshka Coleman}, \bibinfo{person}{Sarina Till}, \bibinfo{person}{Jaydon Farao}, \bibinfo{person}{Londiwe Shandu}, \bibinfo{person}{Nonkululeko Khuzwayo}, \bibinfo{person}{Livhuwani Muthelo}, \bibinfo{person}{Masenyani Mbombi}, \bibinfo{person}{Mamare Bopape}, \bibinfo{person}{Alastair van Heerden}, \bibinfo{person}{Tebogo Mothiba}, {et~al\mbox{.}}} \bibinfo{year}{2023}\natexlab{}.
\newblock \showarticletitle{Reconsidering priorities for digital maternal and child health: community-centered perspectives from South Africa}.
\newblock \bibinfo{journal}{\emph{Proceedings of the ACM on Human-Computer Interaction}} \bibinfo{volume}{7}, \bibinfo{number}{CSCW2} (\bibinfo{year}{2023}), \bibinfo{pages}{1--31}.
\newblock


\bibitem[\protect\citeauthoryear{Corbett and Loukissas}{Corbett and Loukissas}{2019}]%
        {corbett2019engaging}
\bibfield{author}{\bibinfo{person}{Eric Corbett} {and} \bibinfo{person}{Yanni Loukissas}.} \bibinfo{year}{2019}\natexlab{}.
\newblock \showarticletitle{Engaging gentrification as a social justice issue in HCI}. In \bibinfo{booktitle}{\emph{Proceedings of the 2019 chi conference on human factors in computing systems}}. \bibinfo{pages}{1--16}.
\newblock


\bibitem[\protect\citeauthoryear{Correa and Reichmann}{Correa and Reichmann}{1994}]%
        {correa1994population}
\bibfield{author}{\bibinfo{person}{Sonia Correa} {and} \bibinfo{person}{Rebecca~Lynn Reichmann}.} \bibinfo{year}{1994}\natexlab{}.
\newblock \bibinfo{booktitle}{\emph{Population and reproductive rights: Feminist perspectives from the South}}.
\newblock \bibinfo{publisher}{Zed Books}.
\newblock


\bibitem[\protect\citeauthoryear{Critchley, Babayev, Bulun, Clark, Garcia-Grau, Gregersen, Kilcoyne, Kim, Lavender, Marsh, et~al\mbox{.}}{Critchley et~al\mbox{.}}{2020}]%
        {critchley2020menstruation}
\bibfield{author}{\bibinfo{person}{Hilary~OD Critchley}, \bibinfo{person}{Elnur Babayev}, \bibinfo{person}{Serdar~E Bulun}, \bibinfo{person}{Sandy Clark}, \bibinfo{person}{Iolanda Garcia-Grau}, \bibinfo{person}{Peter~K Gregersen}, \bibinfo{person}{Aoife Kilcoyne}, \bibinfo{person}{Ji-Yong~Julie Kim}, \bibinfo{person}{Missy Lavender}, \bibinfo{person}{Erica~E Marsh}, {et~al\mbox{.}}} \bibinfo{year}{2020}\natexlab{}.
\newblock \showarticletitle{Menstruation: science and society}.
\newblock \bibinfo{journal}{\emph{American journal of obstetrics and gynecology}} \bibinfo{volume}{223}, \bibinfo{number}{5} (\bibinfo{year}{2020}), \bibinfo{pages}{624--664}.
\newblock


\bibitem[\protect\citeauthoryear{Davis and Nathan}{Davis and Nathan}{2015}]%
        {davis2015value}
\bibfield{author}{\bibinfo{person}{Janet Davis} {and} \bibinfo{person}{Lisa~P Nathan}.} \bibinfo{year}{2015}\natexlab{}.
\newblock \showarticletitle{Value sensitive design: Applications, adaptations, and critiques}.
\newblock \bibinfo{journal}{\emph{Handbook of ethics, values, and technological design: Sources, theory, values and application domains}} (\bibinfo{year}{2015}), \bibinfo{pages}{11--40}.
\newblock


\bibitem[\protect\citeauthoryear{De~Onis}{De~Onis}{2012}]%
        {de2012looking}
\bibfield{author}{\bibinfo{person}{Kathleen~M De~Onis}.} \bibinfo{year}{2012}\natexlab{}.
\newblock \showarticletitle{“Looking both ways”: Metaphor and the rhetorical alignment of intersectional climate justice and reproductive justice concerns}.
\newblock \bibinfo{journal}{\emph{Environmental Communication: A Journal of Nature and Culture}} \bibinfo{volume}{6}, \bibinfo{number}{3} (\bibinfo{year}{2012}), \bibinfo{pages}{308--327}.
\newblock


\bibitem[\protect\citeauthoryear{Dearden}{Dearden}{2012}]%
        {99dearden2012see}
\bibfield{author}{\bibinfo{person}{Andy Dearden}.} \bibinfo{year}{2012}\natexlab{}.
\newblock \showarticletitle{See no evil?: ethics in an interventionist ICTD}. In \bibinfo{booktitle}{\emph{Proceedings of the fifth international conference on information and communication technologies and development}}. ACM, \bibinfo{pages}{46--55}.
\newblock


\bibitem[\protect\citeauthoryear{Dearden and Tucker}{Dearden and Tucker}{2015}]%
        {100dearden2015ethical}
\bibfield{author}{\bibinfo{person}{Andy Dearden} {and} \bibinfo{person}{William~D Tucker}.} \bibinfo{year}{2015}\natexlab{}.
\newblock \showarticletitle{The ethical limits of bungee research in ICTD}. In \bibinfo{booktitle}{\emph{Technology and Society (ISTAS), 2015 IEEE International Symposium on}}. IEEE, \bibinfo{pages}{1--6}.
\newblock


\bibitem[\protect\citeauthoryear{DeClerque, Tsui, Abul-Ata, and Barcelona}{DeClerque et~al\mbox{.}}{1986}]%
        {declerque1986rumor}
\bibfield{author}{\bibinfo{person}{Julia DeClerque}, \bibinfo{person}{Amy~Ong Tsui}, \bibinfo{person}{Mohammed~Futuah Abul-Ata}, {and} \bibinfo{person}{Delia Barcelona}.} \bibinfo{year}{1986}\natexlab{}.
\newblock \showarticletitle{Rumor, misinformation and oral contraceptive use in Egypt}.
\newblock \bibinfo{journal}{\emph{Social Science \& Medicine}} \bibinfo{volume}{23}, \bibinfo{number}{1} (\bibinfo{year}{1986}), \bibinfo{pages}{83--92}.
\newblock


\bibitem[\protect\citeauthoryear{Dickison}{Dickison}{1973}]%
        {dickison1973abortion}
\bibfield{author}{\bibinfo{person}{Sheila~K Dickison}.} \bibinfo{year}{1973}\natexlab{}.
\newblock \showarticletitle{Abortion in antiquity}.
\newblock \bibinfo{journal}{\emph{Arethusa}} \bibinfo{volume}{6}, \bibinfo{number}{1} (\bibinfo{year}{1973}), \bibinfo{pages}{159--166}.
\newblock


\bibitem[\protect\citeauthoryear{DiSalvo, Sengers, and Brynjarsd{\'o}ttir}{DiSalvo et~al\mbox{.}}{2010}]%
        {97disalvo2010mapping}
\bibfield{author}{\bibinfo{person}{Carl DiSalvo}, \bibinfo{person}{Phoebe Sengers}, {and} \bibinfo{person}{Hr{\"o}nn Brynjarsd{\'o}ttir}.} \bibinfo{year}{2010}\natexlab{}.
\newblock \showarticletitle{Mapping the landscape of sustainable HCI}. In \bibinfo{booktitle}{\emph{Proceedings of the SIGCHI conference on human factors in computing systems}}. ACM, \bibinfo{pages}{1975--1984}.
\newblock


\bibitem[\protect\citeauthoryear{Dodson and Sterling}{Dodson and Sterling}{2012}]%
        {101dodson2012ethics}
\bibfield{author}{\bibinfo{person}{Leslie Dodson} {and} \bibinfo{person}{S~Revi Sterling}.} \bibinfo{year}{2012}\natexlab{}.
\newblock \showarticletitle{Ethics of Participation: Research or Reporting?}
\newblock \bibinfo{journal}{\emph{The Electronic Journal of Information Systems in Developing Countries}} \bibinfo{volume}{50}, \bibinfo{number}{1} (\bibinfo{year}{2012}), \bibinfo{pages}{1--14}.
\newblock


\bibitem[\protect\citeauthoryear{Edwards, Bellotti, Dey, and Newman}{Edwards et~al\mbox{.}}{2003}]%
        {edwards2003challenges}
\bibfield{author}{\bibinfo{person}{W~Keith Edwards}, \bibinfo{person}{Victoria Bellotti}, \bibinfo{person}{Anind~K Dey}, {and} \bibinfo{person}{Mark~W Newman}.} \bibinfo{year}{2003}\natexlab{}.
\newblock \showarticletitle{The challenges of user-centered design and evaluation for infrastructure}. In \bibinfo{booktitle}{\emph{Proceedings of the SIGCHI conference on Human factors in computing systems}}. \bibinfo{pages}{297--304}.
\newblock


\bibitem[\protect\citeauthoryear{El~Dawla}{El~Dawla}{1999}]%
        {el1999political}
\bibfield{author}{\bibinfo{person}{Aida~Seif El~Dawla}.} \bibinfo{year}{1999}\natexlab{}.
\newblock \showarticletitle{The political and legal struggle over female genital mutilation in Egypt: five years since the ICPD}.
\newblock \bibinfo{journal}{\emph{Reproductive health matters}} \bibinfo{volume}{7}, \bibinfo{number}{13} (\bibinfo{year}{1999}), \bibinfo{pages}{128--136}.
\newblock


\bibitem[\protect\citeauthoryear{Epstein, Lee, Kang, Agapie, Schroeder, Pina, Fogarty, Kientz, and Munson}{Epstein et~al\mbox{.}}{2017}]%
        {epstein2017examining}
\bibfield{author}{\bibinfo{person}{Daniel~A Epstein}, \bibinfo{person}{Nicole~B Lee}, \bibinfo{person}{Jennifer~H Kang}, \bibinfo{person}{Elena Agapie}, \bibinfo{person}{Jessica Schroeder}, \bibinfo{person}{Laura~R Pina}, \bibinfo{person}{James Fogarty}, \bibinfo{person}{Julie~A Kientz}, {and} \bibinfo{person}{Sean Munson}.} \bibinfo{year}{2017}\natexlab{}.
\newblock \showarticletitle{Examining menstrual tracking to inform the design of personal informatics tools}. In \bibinfo{booktitle}{\emph{Proceedings of the 2017 CHI conference on human factors in computing systems}}. \bibinfo{pages}{6876--6888}.
\newblock


\bibitem[\protect\citeauthoryear{Eshak}{Eshak}{2020}]%
        {eshak2020myths}
\bibfield{author}{\bibinfo{person}{Ehab Eshak}.} \bibinfo{year}{2020}\natexlab{}.
\newblock \showarticletitle{Myths about modern and traditional contraceptives held by women in Minia, Upper Egypt}.
\newblock \bibinfo{journal}{\emph{East Mediterr Health J}} \bibinfo{volume}{26}, \bibinfo{number}{4} (\bibinfo{year}{2020}), \bibinfo{pages}{417--425}.
\newblock


\bibitem[\protect\citeauthoryear{Figma}{Figma}{2021}]%
        {Figma:online}
\bibfield{author}{\bibinfo{person}{Figma}.} \bibinfo{year}{2021}\natexlab{}.
\newblock \bibinfo{title}{The prototyping tool for teams}.
\newblock \bibinfo{howpublished}{\url{https://www.figma.com/}}.
\newblock


\bibitem[\protect\citeauthoryear{Fox, Asad, Lo, Dimond, Dombrowski, and Bardzell}{Fox et~al\mbox{.}}{2016}]%
        {fox2016exploring}
\bibfield{author}{\bibinfo{person}{Sarah Fox}, \bibinfo{person}{Mariam Asad}, \bibinfo{person}{Katherine Lo}, \bibinfo{person}{Jill~P Dimond}, \bibinfo{person}{Lynn~S Dombrowski}, {and} \bibinfo{person}{Shaowen Bardzell}.} \bibinfo{year}{2016}\natexlab{}.
\newblock \showarticletitle{Exploring social justice, design, and HCI}. In \bibinfo{booktitle}{\emph{Proceedings of the 2016 CHI Conference Extended Abstracts on Human Factors in Computing Systems}}. \bibinfo{pages}{3293--3300}.
\newblock


\bibitem[\protect\citeauthoryear{Fox, Dimond, Irani, Hirsch, Muller, and Bardzell}{Fox et~al\mbox{.}}{2017}]%
        {fox2017social}
\bibfield{author}{\bibinfo{person}{Sarah Fox}, \bibinfo{person}{Jill Dimond}, \bibinfo{person}{Lilly Irani}, \bibinfo{person}{Tad Hirsch}, \bibinfo{person}{Michael Muller}, {and} \bibinfo{person}{Shaowen Bardzell}.} \bibinfo{year}{2017}\natexlab{}.
\newblock \showarticletitle{Social Justice and Design: Power and oppression in collaborative systems}. In \bibinfo{booktitle}{\emph{Companion of the 2017 acm conference on computer supported cooperative work and social computing}}. \bibinfo{pages}{117--122}.
\newblock


\bibitem[\protect\citeauthoryear{Friedman}{Friedman}{1996}]%
        {friedman1996value}
\bibfield{author}{\bibinfo{person}{Batya Friedman}.} \bibinfo{year}{1996}\natexlab{}.
\newblock \showarticletitle{Value-sensitive design}.
\newblock \bibinfo{journal}{\emph{interactions}} \bibinfo{volume}{3}, \bibinfo{number}{6} (\bibinfo{year}{1996}), \bibinfo{pages}{16--23}.
\newblock


\bibitem[\protect\citeauthoryear{Friedman, Kahn, and Borning}{Friedman et~al\mbox{.}}{2002}]%
        {friedman2002value}
\bibfield{author}{\bibinfo{person}{Batya Friedman}, \bibinfo{person}{Peter Kahn}, {and} \bibinfo{person}{Alan Borning}.} \bibinfo{year}{2002}\natexlab{}.
\newblock \showarticletitle{Value sensitive design: Theory and methods}.
\newblock \bibinfo{journal}{\emph{University of Washington technical report}} \bibinfo{number}{2-12} (\bibinfo{year}{2002}).
\newblock


\bibitem[\protect\citeauthoryear{Friedman, Kahn, Borning, and Huldtgren}{Friedman et~al\mbox{.}}{2013}]%
        {friedman2013value}
\bibfield{author}{\bibinfo{person}{Batya Friedman}, \bibinfo{person}{Peter~H Kahn}, \bibinfo{person}{Alan Borning}, {and} \bibinfo{person}{Alina Huldtgren}.} \bibinfo{year}{2013}\natexlab{}.
\newblock \showarticletitle{Value sensitive design and information systems}.
\newblock In \bibinfo{booktitle}{\emph{Early engagement and new technologies: Opening up the laboratory}}. \bibinfo{publisher}{Springer}, \bibinfo{pages}{55--95}.
\newblock


\bibitem[\protect\citeauthoryear{Ghule, Raj, Palaye, Dasgupta, Nair, Saggurti, Battala, and Balaiah}{Ghule et~al\mbox{.}}{2015}]%
        {ghule2015barriers}
\bibfield{author}{\bibinfo{person}{Mohan Ghule}, \bibinfo{person}{Anita Raj}, \bibinfo{person}{Prajakta Palaye}, \bibinfo{person}{Anindita Dasgupta}, \bibinfo{person}{Saritha Nair}, \bibinfo{person}{Niranjan Saggurti}, \bibinfo{person}{Madhusudana Battala}, {and} \bibinfo{person}{Donta Balaiah}.} \bibinfo{year}{2015}\natexlab{}.
\newblock \showarticletitle{Barriers to use contraceptive methods among rural young married couples in Maharashtra, India: qualitative findings}.
\newblock \bibinfo{journal}{\emph{Asian journal of research in social sciences and humanities}} \bibinfo{volume}{5}, \bibinfo{number}{6} (\bibinfo{year}{2015}), \bibinfo{pages}{18}.
\newblock


\bibitem[\protect\citeauthoryear{Hennegan, Shannon, Rubli, Schwab, and Melendez-Torres}{Hennegan et~al\mbox{.}}{2019}]%
        {hennegan2019women}
\bibfield{author}{\bibinfo{person}{Julie Hennegan}, \bibinfo{person}{Alexandra~K Shannon}, \bibinfo{person}{Jennifer Rubli}, \bibinfo{person}{Kellogg~J Schwab}, {and} \bibinfo{person}{GJ Melendez-Torres}.} \bibinfo{year}{2019}\natexlab{}.
\newblock \showarticletitle{Women’s and girls’ experiences of menstruation in low-and middle-income countries: A systematic review and qualitative metasynthesis}.
\newblock \bibinfo{journal}{\emph{PLoS medicine}} \bibinfo{volume}{16}, \bibinfo{number}{5} (\bibinfo{year}{2019}), \bibinfo{pages}{e1002803}.
\newblock


\bibitem[\protect\citeauthoryear{Hessini}{Hessini}{2007}]%
        {hessini2007abortion}
\bibfield{author}{\bibinfo{person}{Leila Hessini}.} \bibinfo{year}{2007}\natexlab{}.
\newblock \showarticletitle{Abortion and Islam: policies and practice in the Middle East and North Africa}.
\newblock \bibinfo{journal}{\emph{Reproductive health matters}} \bibinfo{volume}{15}, \bibinfo{number}{29} (\bibinfo{year}{2007}), \bibinfo{pages}{75--84}.
\newblock


\bibitem[\protect\citeauthoryear{Holtzblatt, Wendell, and Wood}{Holtzblatt et~al\mbox{.}}{2004}]%
        {holtzblatt2004rapid}
\bibfield{author}{\bibinfo{person}{Karen Holtzblatt}, \bibinfo{person}{Jessamyn~Burns Wendell}, {and} \bibinfo{person}{Shelley Wood}.} \bibinfo{year}{2004}\natexlab{}.
\newblock \bibinfo{booktitle}{\emph{Rapid contextual design: a how-to guide to key techniques for user-centered design}}.
\newblock \bibinfo{publisher}{Elsevier}.
\newblock


\bibitem[\protect\citeauthoryear{Homewood, Hedemyr, Fagerberg~Ranten, and Kozel}{Homewood et~al\mbox{.}}{2021}]%
        {homewood2021tracing}
\bibfield{author}{\bibinfo{person}{Sarah Homewood}, \bibinfo{person}{Marika Hedemyr}, \bibinfo{person}{Maja Fagerberg~Ranten}, {and} \bibinfo{person}{Susan Kozel}.} \bibinfo{year}{2021}\natexlab{}.
\newblock \showarticletitle{Tracing conceptions of the body in HCI: From user to more-than-human}. In \bibinfo{booktitle}{\emph{Proceedings of the 2021 CHI Conference on Human Factors in Computing Systems}}. \bibinfo{pages}{1--12}.
\newblock


\bibitem[\protect\citeauthoryear{Hossain, Shohel, Jahan, and Sultana}{Hossain et~al\mbox{.}}{2017}]%
        {hossain2017knowledge}
\bibfield{author}{\bibinfo{person}{Md~Tanvir Hossain}, \bibinfo{person}{Tunvir~Ahamed Shohel}, \bibinfo{person}{Nusrat Jahan}, {and} \bibinfo{person}{Nahida Sultana}.} \bibinfo{year}{2017}\natexlab{}.
\newblock \showarticletitle{Knowledge of female adolescents about reproductive health in South-Western region of Bangladesh}.
\newblock \bibinfo{journal}{\emph{Khulna University Studies}} (\bibinfo{year}{2017}), \bibinfo{pages}{149--161}.
\newblock


\bibitem[\protect\citeauthoryear{Huber and Khan}{Huber and Khan}{1979}]%
        {huber1979contraceptive}
\bibfield{author}{\bibinfo{person}{Douglas~H Huber} {and} \bibinfo{person}{Atiqur~Rahman Khan}.} \bibinfo{year}{1979}\natexlab{}.
\newblock \showarticletitle{Contraceptive distribution in Bangladesh villages: the initial impact}.
\newblock \bibinfo{journal}{\emph{Studies in family planning}} \bibinfo{volume}{10}, \bibinfo{number}{8/9} (\bibinfo{year}{1979}), \bibinfo{pages}{246--253}.
\newblock


\bibitem[\protect\citeauthoryear{Jain and Yammiyavar}{Jain and Yammiyavar}{2015}]%
        {jain2015game}
\bibfield{author}{\bibinfo{person}{Minal Jain} {and} \bibinfo{person}{Pradeep Yammiyavar}.} \bibinfo{year}{2015}\natexlab{}.
\newblock \showarticletitle{Game based learning tool seeking peer support for empowering adolescent girls in rural Assam}. In \bibinfo{booktitle}{\emph{Proceedings of the 14th International Conference on Interaction Design and Children}}. \bibinfo{pages}{275--278}.
\newblock


\bibitem[\protect\citeauthoryear{Jain, Ibtasam, Sharma, Bhattacharya, Tuli, Gamage, Jain, Tulaskar, Chandra, Razaq, et~al\mbox{.}}{Jain et~al\mbox{.}}{2021}]%
        {jain2021margins}
\bibfield{author}{\bibinfo{person}{Pranjal Jain}, \bibinfo{person}{Samia Ibtasam}, \bibinfo{person}{Sumita Sharma}, \bibinfo{person}{Nilavra Bhattacharya}, \bibinfo{person}{Anupriya Tuli}, \bibinfo{person}{Dilrukshi Gamage}, \bibinfo{person}{Dhruv Jain}, \bibinfo{person}{Rucha Tulaskar}, \bibinfo{person}{Priyank Chandra}, \bibinfo{person}{Lubna Razaq}, {et~al\mbox{.}}} \bibinfo{year}{2021}\natexlab{}.
\newblock \showarticletitle{From the Margins to the Centre: Defining New Mission and Vision for HCI Research in South Asia}. In \bibinfo{booktitle}{\emph{Extended Abstracts of the 2021 CHI Conference on Human Factors in Computing Systems}}. \bibinfo{pages}{1--6}.
\newblock


\bibitem[\protect\citeauthoryear{Jaspan, Gray, Robinson, Coovadia, and Bekker}{Jaspan et~al\mbox{.}}{2005}]%
        {jaspan2005scientific}
\bibfield{author}{\bibinfo{person}{Heather~B Jaspan}, \bibinfo{person}{Glenda~E Gray}, \bibinfo{person}{Andrew~KL Robinson}, \bibinfo{person}{Hoosen~M Coovadia}, {and} \bibinfo{person}{Linda-Gail Bekker}.} \bibinfo{year}{2005}\natexlab{}.
\newblock \showarticletitle{Scientific justification for the participation of children and adolescents in HIV-1 vaccine trials in South Africa}.
\newblock \bibinfo{journal}{\emph{South African Medical Journal}} \bibinfo{volume}{95}, \bibinfo{number}{9} (\bibinfo{year}{2005}).
\newblock


\bibitem[\protect\citeauthoryear{Joshi and Schultz}{Joshi and Schultz}{2013}]%
        {joshi2013family}
\bibfield{author}{\bibinfo{person}{Shareen Joshi} {and} \bibinfo{person}{T~Paul Schultz}.} \bibinfo{year}{2013}\natexlab{}.
\newblock \showarticletitle{Family planning and women’s and children’s health: Long-term consequences of an outreach program in Matlab, Bangladesh}.
\newblock \bibinfo{journal}{\emph{Demography}} \bibinfo{volume}{50}, \bibinfo{number}{1} (\bibinfo{year}{2013}), \bibinfo{pages}{149--180}.
\newblock


\bibitem[\protect\citeauthoryear{Kamran}{Kamran}{2023}]%
        {kamran2023decolonizing}
\bibfield{author}{\bibinfo{person}{Areeba Kamran}.} \bibinfo{year}{2023}\natexlab{}.
\newblock \showarticletitle{Decolonizing Artificial Intelligence: Unveiling Biases, Power Dynamics, and Colonial Continuities in AI Systems}.
\newblock \bibinfo{journal}{\emph{RMS journal}} (\bibinfo{year}{2023}).
\newblock


\bibitem[\protect\citeauthoryear{Keinonen}{Keinonen}{2008}]%
        {keinonen2008user}
\bibfield{author}{\bibinfo{person}{Turkka Keinonen}.} \bibinfo{year}{2008}\natexlab{}.
\newblock \showarticletitle{User-centered design and fundamental need}. In \bibinfo{booktitle}{\emph{Proceedings of the 5th Nordic conference on Human-computer interaction: building bridges}}. \bibinfo{pages}{211--219}.
\newblock


\bibitem[\protect\citeauthoryear{Kelty, Panofsky, Currie, Crooks, Erickson, Garcia, Wartenbe, and Wood}{Kelty et~al\mbox{.}}{2015}]%
        {kelty2015seven}
\bibfield{author}{\bibinfo{person}{Christopher Kelty}, \bibinfo{person}{Aaron Panofsky}, \bibinfo{person}{Morgan Currie}, \bibinfo{person}{Roderic Crooks}, \bibinfo{person}{Seth Erickson}, \bibinfo{person}{Patricia Garcia}, \bibinfo{person}{Michael Wartenbe}, {and} \bibinfo{person}{Stacy Wood}.} \bibinfo{year}{2015}\natexlab{}.
\newblock \showarticletitle{Seven dimensions of contemporary participation disentangled}.
\newblock \bibinfo{journal}{\emph{Journal of the Association for Information Science and Technology}} \bibinfo{volume}{66}, \bibinfo{number}{3} (\bibinfo{year}{2015}), \bibinfo{pages}{474--488}.
\newblock


\bibitem[\protect\citeauthoryear{Keyes, Peil, Williams, and Spiel}{Keyes et~al\mbox{.}}{2020}]%
        {keyes2020reimagining}
\bibfield{author}{\bibinfo{person}{Os Keyes}, \bibinfo{person}{Burren Peil}, \bibinfo{person}{Rua~M Williams}, {and} \bibinfo{person}{Katta Spiel}.} \bibinfo{year}{2020}\natexlab{}.
\newblock \showarticletitle{Reimagining (women’s) health: HCI, gender and essentialised embodiment}.
\newblock \bibinfo{journal}{\emph{ACM Transactions on Computer-Human Interaction (TOCHI)}} \bibinfo{volume}{27}, \bibinfo{number}{4} (\bibinfo{year}{2020}), \bibinfo{pages}{1--42}.
\newblock


\bibitem[\protect\citeauthoryear{Klausen}{Klausen}{2019}]%
        {klausen2019reproductive}
\bibfield{author}{\bibinfo{person}{Susanne~M Klausen}.} \bibinfo{year}{2019}\natexlab{}.
\newblock \showarticletitle{Reproductive Health, Fertility Control, and Childbirth in Africa}.
\newblock In \bibinfo{booktitle}{\emph{Oxford Research Encyclopedia of African History}}.
\newblock


\bibitem[\protect\citeauthoryear{Koenig, Jamil, Streatfield, Saha, Al-Sabir, Arifeen, Hill, and Haque}{Koenig et~al\mbox{.}}{2007}]%
        {koenig2007maternal}
\bibfield{author}{\bibinfo{person}{Michael~A Koenig}, \bibinfo{person}{Kanta Jamil}, \bibinfo{person}{Peter~K Streatfield}, \bibinfo{person}{Tulshi Saha}, \bibinfo{person}{Ahmed Al-Sabir}, \bibinfo{person}{Shams~El Arifeen}, \bibinfo{person}{Ken Hill}, {and} \bibinfo{person}{Yasmin Haque}.} \bibinfo{year}{2007}\natexlab{}.
\newblock \showarticletitle{Maternal health and care-seeking behavior in Bangladesh: findings from a national survey}.
\newblock \bibinfo{journal}{\emph{International family planning perspectives}} (\bibinfo{year}{2007}), \bibinfo{pages}{75--82}.
\newblock


\bibitem[\protect\citeauthoryear{Kumar and Anderson}{Kumar and Anderson}{2015}]%
        {kumar2015mobile}
\bibfield{author}{\bibinfo{person}{Neha Kumar} {and} \bibinfo{person}{Richard~J Anderson}.} \bibinfo{year}{2015}\natexlab{}.
\newblock \showarticletitle{Mobile phones for maternal health in rural India}. In \bibinfo{booktitle}{\emph{Proceedings of the 33rd annual acm conference on human factors in computing systems}}. \bibinfo{pages}{427--436}.
\newblock


\bibitem[\protect\citeauthoryear{Kumar and Karusala}{Kumar and Karusala}{2021}]%
        {kumar2021braving}
\bibfield{author}{\bibinfo{person}{Neha Kumar} {and} \bibinfo{person}{Naveena Karusala}.} \bibinfo{year}{2021}\natexlab{}.
\newblock \showarticletitle{Braving citational justice in human-computer interaction}. In \bibinfo{booktitle}{\emph{Extended Abstracts of the 2021 CHI Conference on Human Factors in Computing Systems}}. \bibinfo{pages}{1--9}.
\newblock


\bibitem[\protect\citeauthoryear{Kumar, Karusala, Ismail, and Tuli}{Kumar et~al\mbox{.}}{2020}]%
        {kumar2020taking}
\bibfield{author}{\bibinfo{person}{Neha Kumar}, \bibinfo{person}{Naveena Karusala}, \bibinfo{person}{Azra Ismail}, {and} \bibinfo{person}{Anupriya Tuli}.} \bibinfo{year}{2020}\natexlab{}.
\newblock \showarticletitle{Taking the long, holistic, and intersectional view to women’s wellbeing}.
\newblock \bibinfo{journal}{\emph{ACM Transactions on Computer-Human Interaction (TOCHI)}} \bibinfo{volume}{27}, \bibinfo{number}{4} (\bibinfo{year}{2020}), \bibinfo{pages}{1--32}.
\newblock


\bibitem[\protect\citeauthoryear{Kumar, Perrier, Desmond, Israel-Ballard, Kumar, Mahapatra, Mishra, Agarwal, Gandhi, Lal, et~al\mbox{.}}{Kumar et~al\mbox{.}}{2015}]%
        {kumar2015projecting}
\bibfield{author}{\bibinfo{person}{Neha Kumar}, \bibinfo{person}{Trevor Perrier}, \bibinfo{person}{Michelle Desmond}, \bibinfo{person}{Kiersten Israel-Ballard}, \bibinfo{person}{Vikrant Kumar}, \bibinfo{person}{Sudip Mahapatra}, \bibinfo{person}{Anil Mishra}, \bibinfo{person}{Shreya Agarwal}, \bibinfo{person}{Rikin Gandhi}, \bibinfo{person}{Pallavi Lal}, {et~al\mbox{.}}} \bibinfo{year}{2015}\natexlab{}.
\newblock \showarticletitle{Projecting health: community-led video education for maternal health}. In \bibinfo{booktitle}{\emph{Proceedings of the Seventh International Conference on Information and Communication Technologies and Development}}. ACM, \bibinfo{pages}{17}.
\newblock


\bibitem[\protect\citeauthoryear{Kundu, Kabir, and Islam}{Kundu et~al\mbox{.}}{2020}]%
        {kundu2020evaluating}
\bibfield{author}{\bibinfo{person}{Shusmoy Kundu}, \bibinfo{person}{Anusha Kabir}, {and} \bibinfo{person}{Muhammad~Nazrul Islam}.} \bibinfo{year}{2020}\natexlab{}.
\newblock \showarticletitle{Evaluating usability of pregnancy tracker applications in Bangladesh: a heuristic and semiotic evaluation}. In \bibinfo{booktitle}{\emph{2020 IEEE 8th R10 Humanitarian Technology Conference (R10-HTC)}}. IEEE, \bibinfo{pages}{1--6}.
\newblock


\bibitem[\protect\citeauthoryear{Lazar, Su, Bardzell, and Bardzell}{Lazar et~al\mbox{.}}{2019}]%
        {lazar2019parting}
\bibfield{author}{\bibinfo{person}{Amanda Lazar}, \bibinfo{person}{Norman~Makoto Su}, \bibinfo{person}{Jeffrey Bardzell}, {and} \bibinfo{person}{Shaowen Bardzell}.} \bibinfo{year}{2019}\natexlab{}.
\newblock \showarticletitle{Parting the Red Sea: sociotechnical systems and lived experiences of menopause}. In \bibinfo{booktitle}{\emph{Proceedings of the 2019 CHI conference on human factors in computing systems}}. \bibinfo{pages}{1--16}.
\newblock


\bibitem[\protect\citeauthoryear{Lazem, Giglitto, Nkwo, Mthoko, Upani, and Peters}{Lazem et~al\mbox{.}}{2021}]%
        {lazem2021challenges}
\bibfield{author}{\bibinfo{person}{Shaimaa Lazem}, \bibinfo{person}{Danilo Giglitto}, \bibinfo{person}{Makuochi~Samuel Nkwo}, \bibinfo{person}{Hafeni Mthoko}, \bibinfo{person}{Jessica Upani}, {and} \bibinfo{person}{Anicia Peters}.} \bibinfo{year}{2021}\natexlab{}.
\newblock \showarticletitle{Challenges and paradoxes in decolonising HCI: A critical discussion}.
\newblock \bibinfo{journal}{\emph{Computer Supported Cooperative Work (CSCW)}} (\bibinfo{year}{2021}), \bibinfo{pages}{1--38}.
\newblock


\bibitem[\protect\citeauthoryear{Le~Dantec}{Le~Dantec}{2012}]%
        {le2012participation}
\bibfield{author}{\bibinfo{person}{Christopher Le~Dantec}.} \bibinfo{year}{2012}\natexlab{}.
\newblock \showarticletitle{Participation and publics: supporting community engagement}. In \bibinfo{booktitle}{\emph{Proceedings of the SIGCHI Conference on Human Factors in Computing Systems}}. \bibinfo{pages}{1351--1360}.
\newblock


\bibitem[\protect\citeauthoryear{Lyu, Cai, Callis, Cotter, and Carroll}{Lyu et~al\mbox{.}}{2024}]%
        {lyu2024got}
\bibfield{author}{\bibinfo{person}{Yao Lyu}, \bibinfo{person}{Jie Cai}, \bibinfo{person}{Anisa Callis}, \bibinfo{person}{Kelley Cotter}, {and} \bibinfo{person}{John~M Carroll}.} \bibinfo{year}{2024}\natexlab{}.
\newblock \showarticletitle{" I Got Flagged for Supposed Bullying, Even Though It Was in Response to Someone Harassing Me About My Disability.": A Study of Blind TikTokers’ Content Moderation Experiences}. In \bibinfo{booktitle}{\emph{Proceedings of the CHI Conference on Human Factors in Computing Systems}}. \bibinfo{pages}{1--15}.
\newblock


\bibitem[\protect\citeauthoryear{Mahar, Zhang, and Karger}{Mahar et~al\mbox{.}}{2018}]%
        {mahar2018squadbox}
\bibfield{author}{\bibinfo{person}{Kaitlin Mahar}, \bibinfo{person}{Amy~X Zhang}, {and} \bibinfo{person}{David Karger}.} \bibinfo{year}{2018}\natexlab{}.
\newblock \showarticletitle{Squadbox: A tool to combat email harassment using friendsourced moderation}. In \bibinfo{booktitle}{\emph{Proceedings of the 2018 CHI Conference on Human Factors in Computing Systems}}. \bibinfo{pages}{1--13}.
\newblock


\bibitem[\protect\citeauthoryear{Mahbub, Islam, Surem, Arefin, Shabab, and Islam}{Mahbub et~al\mbox{.}}{2024}]%
        {mahbub2024gorbhokotha}
\bibfield{author}{\bibinfo{person}{Sarah~Binte Mahbub}, \bibinfo{person}{Nafis Islam}, \bibinfo{person}{Mehedi~Hasan Surem}, \bibinfo{person}{Ayatullah Arefin}, \bibinfo{person}{Md~Rashid Shabab}, {and} \bibinfo{person}{Ashraful Islam}.} \bibinfo{year}{2024}\natexlab{}.
\newblock \showarticletitle{GorbhoKotha: a mHealth App for Maternal Support in Bangladesh Utilizing User-Centered Design Principles}. In \bibinfo{booktitle}{\emph{2024 International Congress on Human-Computer Interaction, Optimization and Robotic Applications (HORA)}}. IEEE, \bibinfo{pages}{1--6}.
\newblock


\bibitem[\protect\citeauthoryear{Mao, Vredenburg, Smith, and Carey}{Mao et~al\mbox{.}}{2005}]%
        {mao2005state}
\bibfield{author}{\bibinfo{person}{Ji-Ye Mao}, \bibinfo{person}{Karel Vredenburg}, \bibinfo{person}{Paul~W Smith}, {and} \bibinfo{person}{Tom Carey}.} \bibinfo{year}{2005}\natexlab{}.
\newblock \showarticletitle{The state of user-centered design practice}.
\newblock \bibinfo{journal}{\emph{Commun. ACM}} \bibinfo{volume}{48}, \bibinfo{number}{3} (\bibinfo{year}{2005}), \bibinfo{pages}{105--109}.
\newblock


\bibitem[\protect\citeauthoryear{Marks}{Marks}{2003}]%
        {marks2003cage}
\bibfield{author}{\bibinfo{person}{Lara Marks}.} \bibinfo{year}{2003}\natexlab{}.
\newblock \showarticletitle{A ‘Cage’of Ovulating Females”: The History of the Early Oral Contraceptive Pill Clinical Trials, 1950--1959.”}.
\newblock \bibinfo{journal}{\emph{Molecularizing Biology and Medicine: New Practices and Alliances, 1920s to 1970s}} (\bibinfo{year}{2003}), \bibinfo{pages}{208}.
\newblock


\bibitem[\protect\citeauthoryear{McDonald, Zhao, Liu, and Rivera}{McDonald et~al\mbox{.}}{2018}]%
        {mcdonald2018maxifab}
\bibfield{author}{\bibinfo{person}{Joselyn McDonald}, \bibinfo{person}{Siyan Zhao}, \bibinfo{person}{Jen Liu}, {and} \bibinfo{person}{Michael~L Rivera}.} \bibinfo{year}{2018}\natexlab{}.
\newblock \showarticletitle{MaxiFab: Applied fabrication to advance period technologies}. In \bibinfo{booktitle}{\emph{Proceedings of the 2018 ACM Conference Companion Publication on Designing Interactive Systems}}. \bibinfo{pages}{13--19}.
\newblock


\bibitem[\protect\citeauthoryear{Mehrnezhad and Almeida}{Mehrnezhad and Almeida}{2023}]%
        {mehrnezhad2023my}
\bibfield{author}{\bibinfo{person}{Maryam Mehrnezhad} {and} \bibinfo{person}{Teresa Almeida}.} \bibinfo{year}{2023}\natexlab{}.
\newblock \showarticletitle{“My sex-related data is more sensitive than my financial data and I want the same level of security and privacy$\backslash$": User Risk Perceptions and Protective Actions in Female-oriented Technologies}. In \bibinfo{booktitle}{\emph{Proceedings of the 2023 European Symposium on Usable Security}}. \bibinfo{pages}{1--14}.
\newblock


\bibitem[\protect\citeauthoryear{Michie, Balaam, McCarthy, Osadchiy, and Morrissey}{Michie et~al\mbox{.}}{2018}]%
        {michie2018her}
\bibfield{author}{\bibinfo{person}{Lydia Michie}, \bibinfo{person}{Madeline Balaam}, \bibinfo{person}{John McCarthy}, \bibinfo{person}{Timur Osadchiy}, {and} \bibinfo{person}{Kellie Morrissey}.} \bibinfo{year}{2018}\natexlab{}.
\newblock \showarticletitle{From her story, to our story: Digital storytelling as public engagement around abortion rights advocacy in Ireland}. In \bibinfo{booktitle}{\emph{Proceedings of the 2018 CHI Conference on Human Factors in Computing Systems}}. \bibinfo{pages}{1--15}.
\newblock


\bibitem[\protect\citeauthoryear{Mohamed, Png, and Isaac}{Mohamed et~al\mbox{.}}{2020}]%
        {mohamed2020decolonial}
\bibfield{author}{\bibinfo{person}{Shakir Mohamed}, \bibinfo{person}{Marie-Therese Png}, {and} \bibinfo{person}{William Isaac}.} \bibinfo{year}{2020}\natexlab{}.
\newblock \showarticletitle{Decolonial AI: Decolonial theory as sociotechnical foresight in artificial intelligence}.
\newblock \bibinfo{journal}{\emph{Philosophy \& Technology}}  \bibinfo{volume}{33} (\bibinfo{year}{2020}), \bibinfo{pages}{659--684}.
\newblock


\bibitem[\protect\citeauthoryear{Mugerwa, Kaleebu, Mugyenyi, Katongole-Mbidde, Hom, Byaruhanga, Salata, and Ellner}{Mugerwa et~al\mbox{.}}{2002}]%
        {mugerwa2002first}
\bibfield{author}{\bibinfo{person}{Roy~D Mugerwa}, \bibinfo{person}{Pontiano Kaleebu}, \bibinfo{person}{Peter Mugyenyi}, \bibinfo{person}{Edward Katongole-Mbidde}, \bibinfo{person}{David~L Hom}, \bibinfo{person}{Rose Byaruhanga}, \bibinfo{person}{Robert~A Salata}, {and} \bibinfo{person}{Jerrold~J Ellner}.} \bibinfo{year}{2002}\natexlab{}.
\newblock \showarticletitle{First trial of the HIV-1 vaccine in Africa: Ugandan experience}.
\newblock \bibinfo{journal}{\emph{Bmj}} \bibinfo{volume}{324}, \bibinfo{number}{7331} (\bibinfo{year}{2002}), \bibinfo{pages}{226--229}.
\newblock


\bibitem[\protect\citeauthoryear{Murphy and Nagel}{Murphy and Nagel}{2002}]%
        {murphy2002myth}
\bibfield{author}{\bibinfo{person}{Liam Murphy} {and} \bibinfo{person}{Thomas Nagel}.} \bibinfo{year}{2002}\natexlab{}.
\newblock \bibinfo{booktitle}{\emph{The Myth of Ownership: Taxes and Justice}}.
\newblock \bibinfo{publisher}{Oxford University Press}, \bibinfo{address}{Oxford, UK}.
\newblock


\bibitem[\protect\citeauthoryear{Mustafa, Batool, Fatima, Nawaz, Toyama, and Raza}{Mustafa et~al\mbox{.}}{2020}]%
        {mustafa2020patriarchy}
\bibfield{author}{\bibinfo{person}{Maryam Mustafa}, \bibinfo{person}{Amna Batool}, \bibinfo{person}{Beenish Fatima}, \bibinfo{person}{Fareeda Nawaz}, \bibinfo{person}{Kentaro Toyama}, {and} \bibinfo{person}{Agha~Ali Raza}.} \bibinfo{year}{2020}\natexlab{}.
\newblock \showarticletitle{Patriarchy, maternal health and spiritual healing: Designing maternal health interventions in Pakistan}. In \bibinfo{booktitle}{\emph{Proceedings of the 2020 CHI Conference on Human Factors in Computing Systems}}. \bibinfo{pages}{1--13}.
\newblock


\bibitem[\protect\citeauthoryear{Mustafa, Zaman, Ahmad, Batool, Ghazali, and Ahmed}{Mustafa et~al\mbox{.}}{2021}]%
        {mustafa2021religion}
\bibfield{author}{\bibinfo{person}{Maryam Mustafa}, \bibinfo{person}{Kimia~Tuz Zaman}, \bibinfo{person}{Tallal Ahmad}, \bibinfo{person}{Amna Batool}, \bibinfo{person}{Masitah Ghazali}, {and} \bibinfo{person}{Nova Ahmed}.} \bibinfo{year}{2021}\natexlab{}.
\newblock \showarticletitle{Religion and women’s intimate health: towards an inclusive approach to healthcare}. In \bibinfo{booktitle}{\emph{Proceedings of the 2021 CHI Conference on Human Factors in Computing Systems}}. \bibinfo{pages}{1--13}.
\newblock


\bibitem[\protect\citeauthoryear{Ng, Bardzell, and Bardzell}{Ng et~al\mbox{.}}{2020}]%
        {ng2020menstruating}
\bibfield{author}{\bibinfo{person}{Sarah Ng}, \bibinfo{person}{Shaowen Bardzell}, {and} \bibinfo{person}{Jeffrey Bardzell}.} \bibinfo{year}{2020}\natexlab{}.
\newblock \showarticletitle{The menstruating entrepreneur kickstarting a new politics of women's health}.
\newblock \bibinfo{journal}{\emph{ACM Transactions on Computer-Human Interaction (TOCHI)}} \bibinfo{volume}{27}, \bibinfo{number}{4} (\bibinfo{year}{2020}), \bibinfo{pages}{1--25}.
\newblock


\bibitem[\protect\citeauthoryear{Norman}{Norman}{2005}]%
        {norman2005human}
\bibfield{author}{\bibinfo{person}{Donald~A Norman}.} \bibinfo{year}{2005}\natexlab{}.
\newblock \showarticletitle{Human-centered design considered harmful}.
\newblock \bibinfo{journal}{\emph{interactions}} \bibinfo{volume}{12}, \bibinfo{number}{4} (\bibinfo{year}{2005}), \bibinfo{pages}{14--19}.
\newblock


\bibitem[\protect\citeauthoryear{Nouwen, Van~Mechelen, and Zaman}{Nouwen et~al\mbox{.}}{2015}]%
        {nouwen2015value}
\bibfield{author}{\bibinfo{person}{Marije Nouwen}, \bibinfo{person}{Maarten Van~Mechelen}, {and} \bibinfo{person}{Bieke Zaman}.} \bibinfo{year}{2015}\natexlab{}.
\newblock \showarticletitle{A value sensitive design approach to parental software for young children}. In \bibinfo{booktitle}{\emph{Proceedings of the 14th International Conference on Interaction Design and Children}}. \bibinfo{pages}{363--366}.
\newblock


\bibitem[\protect\citeauthoryear{Nyblade, Singh, Ashburn, Brady, and Olenja}{Nyblade et~al\mbox{.}}{2011}]%
        {nyblade2011once}
\bibfield{author}{\bibinfo{person}{Laura Nyblade}, \bibinfo{person}{Sagri Singh}, \bibinfo{person}{Kim Ashburn}, \bibinfo{person}{Laura Brady}, {and} \bibinfo{person}{Joyce Olenja}.} \bibinfo{year}{2011}\natexlab{}.
\newblock \showarticletitle{“Once I begin to participate, people will run away from me”: Understanding stigma as a barrier to HIV vaccine research participation in Kenya}.
\newblock \bibinfo{journal}{\emph{Vaccine}} \bibinfo{volume}{29}, \bibinfo{number}{48} (\bibinfo{year}{2011}), \bibinfo{pages}{8924--8928}.
\newblock


\bibitem[\protect\citeauthoryear{of~Health and Welfare}{of~Health and Welfare}{2023}]%
        {dhsprogr7:online}
\bibfield{author}{\bibinfo{person}{Ministry of Health} {and} \bibinfo{person}{Family Welfare}.} \bibinfo{year}{2023}\natexlab{}.
\newblock \bibinfo{title}{Bangladesh Demographic and Health Survey 2022}.
\newblock \bibinfo{howpublished}{\url{https://dhsprogram.com/pubs/pdf/PR148/PR148.pdf}}.
\newblock


\bibitem[\protect\citeauthoryear{{\"O}hlund}{{\"O}hlund}{2023}]%
        {ohlund2023social}
\bibfield{author}{\bibinfo{person}{Linnea {\"O}hlund}.} \bibinfo{year}{2023}\natexlab{}.
\newblock \showarticletitle{A Social Justice-Oriented Perspective on Older Adults Technology Use in HCI: Three Opportunities for Societal Inclusion}. In \bibinfo{booktitle}{\emph{International Conference on Human-Computer Interaction}}. Springer, \bibinfo{pages}{519--532}.
\newblock


\bibitem[\protect\citeauthoryear{Perrier, Dell, DeRenzi, Anderson, Kinuthia, Unger, and John-Stewart}{Perrier et~al\mbox{.}}{2015}]%
        {perrier2015engaging}
\bibfield{author}{\bibinfo{person}{Trevor Perrier}, \bibinfo{person}{Nicola Dell}, \bibinfo{person}{Brian DeRenzi}, \bibinfo{person}{Richard Anderson}, \bibinfo{person}{John Kinuthia}, \bibinfo{person}{Jennifer Unger}, {and} \bibinfo{person}{Grace John-Stewart}.} \bibinfo{year}{2015}\natexlab{}.
\newblock \showarticletitle{Engaging pregnant women in Kenya with a hybrid computer-human SMS communication system}. In \bibinfo{booktitle}{\emph{Proceedings of the 33rd Annual ACM Conference on Human Factors in Computing Systems}}. ACM, \bibinfo{pages}{1429--1438}.
\newblock


\bibitem[\protect\citeauthoryear{Petti, Polage, Quinn, Ronald, and Sande}{Petti et~al\mbox{.}}{2006}]%
        {54petti2006laboratory}
\bibfield{author}{\bibinfo{person}{Cathy~A Petti}, \bibinfo{person}{Christopher~R Polage}, \bibinfo{person}{Thomas~C Quinn}, \bibinfo{person}{Allan~R Ronald}, {and} \bibinfo{person}{Merle~A Sande}.} \bibinfo{year}{2006}\natexlab{}.
\newblock \showarticletitle{Laboratory medicine in Africa: a barrier to effective health care}.
\newblock \bibinfo{journal}{\emph{Clinical Infectious Diseases}} \bibinfo{volume}{42}, \bibinfo{number}{3} (\bibinfo{year}{2006}), \bibinfo{pages}{377--382}.
\newblock


\bibitem[\protect\citeauthoryear{Price}{Price}{2018}]%
        {price2018queering}
\bibfield{author}{\bibinfo{person}{Kimala Price}.} \bibinfo{year}{2018}\natexlab{}.
\newblock \showarticletitle{Queering reproductive justice in the Trump era: A note on political intersectionality}.
\newblock \bibinfo{journal}{\emph{Politics \& Gender}} \bibinfo{volume}{14}, \bibinfo{number}{4} (\bibinfo{year}{2018}), \bibinfo{pages}{581--601}.
\newblock


\bibitem[\protect\citeauthoryear{Rahman}{Rahman}{2024}]%
        {Sharifa68:online}
\bibfield{author}{\bibinfo{person}{Kazi~Nafia Rahman}.} \bibinfo{year}{2024}\natexlab{}.
\newblock \bibinfo{title}{Sharifa’s Story: debate erupts in Bangladesh over textbook lesson on third gender people}.
\newblock \bibinfo{howpublished}{\url{https://bdnews24.com/society/rod076voig}}.
\newblock


\bibitem[\protect\citeauthoryear{Rahman, Haider, Curtis, and Lance}{Rahman et~al\mbox{.}}{2016}]%
        {rahman2016mayer}
\bibfield{author}{\bibinfo{person}{Mizanur Rahman}, \bibinfo{person}{M~Moinuddin Haider}, \bibinfo{person}{Sian~L Curtis}, {and} \bibinfo{person}{Peter~M Lance}.} \bibinfo{year}{2016}\natexlab{}.
\newblock \showarticletitle{The Mayer Hashi large-scale program to increase use of long-acting reversible contraceptives and permanent methods in Bangladesh: explaining the disappointing results. An outcome and process evaluation}.
\newblock \bibinfo{journal}{\emph{Global Health: Science and Practice}} \bibinfo{volume}{4}, \bibinfo{number}{Supplement 2} (\bibinfo{year}{2016}), \bibinfo{pages}{S122--S139}.
\newblock


\bibitem[\protect\citeauthoryear{Rahman, Mosley, Khan, Chowdhury, and Chakraborty}{Rahman et~al\mbox{.}}{1980}]%
        {rahman1980contraceptive}
\bibfield{author}{\bibinfo{person}{Makhlisur Rahman}, \bibinfo{person}{Wiley~Henry Mosley}, \bibinfo{person}{Atiqur~Rahman Khan}, \bibinfo{person}{AI Chowdhury}, {and} \bibinfo{person}{Jyotsnamoy Chakraborty}.} \bibinfo{year}{1980}\natexlab{}.
\newblock \showarticletitle{Contraceptive distribution in Bangladesh: some lessons learned}.
\newblock \bibinfo{journal}{\emph{Studies in family planning}} \bibinfo{volume}{11}, \bibinfo{number}{6} (\bibinfo{year}{1980}), \bibinfo{pages}{191--201}.
\newblock


\bibitem[\protect\citeauthoryear{Rastogi}{Rastogi}{2010}]%
        {95rastogi2010building}
\bibfield{author}{\bibinfo{person}{Sanjeev Rastogi}.} \bibinfo{year}{2010}\natexlab{}.
\newblock \showarticletitle{Building bridges between Ayurveda and modern science}.
\newblock \bibinfo{journal}{\emph{International journal of Ayurveda research}} \bibinfo{volume}{1}, \bibinfo{number}{1} (\bibinfo{year}{2010}), \bibinfo{pages}{41}.
\newblock


\bibitem[\protect\citeauthoryear{Reiman}{Reiman}{1999}]%
        {reiman1999abortion}
\bibfield{author}{\bibinfo{person}{Jeffrey~H Reiman}.} \bibinfo{year}{1999}\natexlab{}.
\newblock \bibinfo{booktitle}{\emph{Abortion and the ways we value human life}}.
\newblock \bibinfo{publisher}{Rowman \& Littlefield}.
\newblock


\bibitem[\protect\citeauthoryear{Rescher}{Rescher}{1966}]%
        {Rescher66dis}
\bibfield{author}{\bibinfo{person}{Nicholas Rescher}.} \bibinfo{year}{1966}\natexlab{}.
\newblock \bibinfo{booktitle}{\emph{Distributive Justice a Constructive Critique of the Utilitarian Theory of Distribution}}.
\newblock \bibinfo{publisher}{University Press of America}.
\newblock


\bibitem[\protect\citeauthoryear{Rifat, Toriq, and Ahmed}{Rifat et~al\mbox{.}}{2020}]%
        {rifat2020religion}
\bibfield{author}{\bibinfo{person}{Mohammad~Rashidujjaman Rifat}, \bibinfo{person}{Toha Toriq}, {and} \bibinfo{person}{Syed~Ishtiaque Ahmed}.} \bibinfo{year}{2020}\natexlab{}.
\newblock \showarticletitle{Religion and Sustainability: Lessons of Sustainable Computing from Islamic Religious Communities}.
\newblock \bibinfo{journal}{\emph{Proceedings of the ACM on Human-Computer Interaction}} \bibinfo{volume}{4}, \bibinfo{number}{CSCW2} (\bibinfo{year}{2020}), \bibinfo{pages}{1--32}.
\newblock


\bibitem[\protect\citeauthoryear{Ross and Solinger}{Ross and Solinger}{2017}]%
        {ross2017reproductive}
\bibfield{author}{\bibinfo{person}{Loretta Ross} {and} \bibinfo{person}{Rickie Solinger}.} \bibinfo{year}{2017}\natexlab{}.
\newblock \bibinfo{booktitle}{\emph{Reproductive justice: An introduction}}. Vol.~\bibinfo{volume}{1}.
\newblock \bibinfo{publisher}{Univ of California Press}.
\newblock


\bibitem[\protect\citeauthoryear{{Rural Reconstruction Foundation}}{{Rural Reconstruction Foundation}}{2017}]%
        {34RRF}
\bibfield{author}{\bibinfo{person}{{Rural Reconstruction Foundation}}.} \bibinfo{year}{2017}\natexlab{}.
\newblock \bibinfo{howpublished}{\url{http://www.rrf-bd.org/}}.
\newblock


\bibitem[\protect\citeauthoryear{Sarder, Alauddin, and Ahammed}{Sarder et~al\mbox{.}}{2020}]%
        {sarder2020determinants}
\bibfield{author}{\bibinfo{person}{Md~Alamgir Sarder}, \bibinfo{person}{Sharlene Alauddin}, {and} \bibinfo{person}{Benojir Ahammed}.} \bibinfo{year}{2020}\natexlab{}.
\newblock \showarticletitle{Determinants of teenage marital pregnancy among bangladeshi women: An analysis by the cox proportional hazard model}.
\newblock \bibinfo{journal}{\emph{Asian Journal of Social Health and Behavior}} \bibinfo{volume}{3}, \bibinfo{number}{4} (\bibinfo{year}{2020}), \bibinfo{pages}{137--143}.
\newblock


\bibitem[\protect\citeauthoryear{Sayem and Nury}{Sayem and Nury}{2011}]%
        {sayem2011factors}
\bibfield{author}{\bibinfo{person}{Amir~M Sayem} {and} \bibinfo{person}{Abu Taher~MS Nury}.} \bibinfo{year}{2011}\natexlab{}.
\newblock \showarticletitle{Factors associated with teenage marital pregnancy among Bangladeshi women}.
\newblock \bibinfo{journal}{\emph{Reproductive health}}  \bibinfo{volume}{8} (\bibinfo{year}{2011}), \bibinfo{pages}{1--6}.
\newblock


\bibitem[\protect\citeauthoryear{Schmidtz, Goodin, Goodin, et~al\mbox{.}}{Schmidtz et~al\mbox{.}}{1998}]%
        {schmidtz1998social}
\bibfield{author}{\bibinfo{person}{David Schmidtz}, \bibinfo{person}{Robert~E Goodin}, \bibinfo{person}{Robert~Edward Goodin}, {et~al\mbox{.}}} \bibinfo{year}{1998}\natexlab{}.
\newblock \bibinfo{booktitle}{\emph{Social Welfare and Individual Responsibility}}.
\newblock \bibinfo{publisher}{Cambridge University Press}.
\newblock


\bibitem[\protect\citeauthoryear{Schroth}{Schroth}{2008}]%
        {schroth2008distributive}
\bibfield{author}{\bibinfo{person}{J{\"o}rg Schroth}.} \bibinfo{year}{2008}\natexlab{}.
\newblock \showarticletitle{Distributive justice and welfarism in utilitarianism}.
\newblock \bibinfo{journal}{\emph{Inquiry}} \bibinfo{volume}{51}, \bibinfo{number}{2} (\bibinfo{year}{2008}), \bibinfo{pages}{123--146}.
\newblock


\bibitem[\protect\citeauthoryear{Scott, Marcu, Anderson, Newman, and Schoenebeck}{Scott et~al\mbox{.}}{2023}]%
        {scott2023trauma}
\bibfield{author}{\bibinfo{person}{Carol~F Scott}, \bibinfo{person}{Gabriela Marcu}, \bibinfo{person}{Riana~Elyse Anderson}, \bibinfo{person}{Mark~W Newman}, {and} \bibinfo{person}{Sarita Schoenebeck}.} \bibinfo{year}{2023}\natexlab{}.
\newblock \showarticletitle{Trauma-informed social media: Towards solutions for reducing and healing online harm}. In \bibinfo{booktitle}{\emph{Proceedings of the 2023 CHI Conference on Human Factors in Computing Systems}}. \bibinfo{pages}{1--20}.
\newblock


\bibitem[\protect\citeauthoryear{Sen, Williams, Williams, et~al\mbox{.}}{Sen et~al\mbox{.}}{1982}]%
        {sen1982utilitarianism}
\bibfield{author}{\bibinfo{person}{Amartya Sen}, \bibinfo{person}{Bernard Arthur~Owen Williams}, \bibinfo{person}{Bernard Williams}, {et~al\mbox{.}}} \bibinfo{year}{1982}\natexlab{}.
\newblock \bibinfo{booktitle}{\emph{Utilitarianism and Beyond}}.
\newblock \bibinfo{publisher}{Cambridge University Press}.
\newblock


\bibitem[\protect\citeauthoryear{Shen, Wan, and Li}{Shen et~al\mbox{.}}{2023}]%
        {shen2023human}
\bibfield{author}{\bibinfo{person}{Pengyi Shen}, \bibinfo{person}{Demin Wan}, {and} \bibinfo{person}{Jinxiong Li}.} \bibinfo{year}{2023}\natexlab{}.
\newblock \showarticletitle{How human--computer interaction perception affects consumer well-being in the context of online retail: From the perspective of autonomy}.
\newblock \bibinfo{journal}{\emph{Nankai Business Review International}} \bibinfo{volume}{14}, \bibinfo{number}{1} (\bibinfo{year}{2023}), \bibinfo{pages}{102--127}.
\newblock


\bibitem[\protect\citeauthoryear{Shirungu}{Shirungu}{2015}]%
        {bidwell2015preserving}
\bibfield{author}{\bibinfo{person}{Michael Shirungu}.} \bibinfo{year}{2015}\natexlab{}.
\newblock \showarticletitle{Preserving `Mpoyetu' (Our Culture), Contestation and Negotiation of Likuki\/Shipumuna Ritual in Kavango Northeast of Namibia}.
\newblock In \bibinfo{booktitle}{\emph{At the Intersection of Indigenous and Traditional Knowledge and Technology Design}}, \bibfield{editor}{\bibinfo{person}{NJ~Bidwell} {and} \bibinfo{person}{H~Winschiers-Theophilus}} (Eds.). \bibinfo{publisher}{Informing Science Press}, \bibinfo{pages}{273--282}.
\newblock


\bibitem[\protect\citeauthoryear{Silberman, Nathan, Knowles, Bendor, Clear, H{\aa}kansson, Dillahunt, and Mankoff}{Silberman et~al\mbox{.}}{2014}]%
        {98silberman2014next}
\bibfield{author}{\bibinfo{person}{M Silberman}, \bibinfo{person}{Lisa Nathan}, \bibinfo{person}{Bran Knowles}, \bibinfo{person}{Roy Bendor}, \bibinfo{person}{Adrian Clear}, \bibinfo{person}{Maria H{\aa}kansson}, \bibinfo{person}{Tawanna Dillahunt}, {and} \bibinfo{person}{Jennifer Mankoff}.} \bibinfo{year}{2014}\natexlab{}.
\newblock \showarticletitle{Next steps for sustainable HCI}.
\newblock \bibinfo{journal}{\emph{interactions}} \bibinfo{volume}{21}, \bibinfo{number}{5} (\bibinfo{year}{2014}), \bibinfo{pages}{66--69}.
\newblock


\bibitem[\protect\citeauthoryear{Singh}{Singh}{2010}]%
        {94singh2010exploring}
\bibfield{author}{\bibinfo{person}{Ram~Harsh Singh}.} \bibinfo{year}{2010}\natexlab{}.
\newblock \showarticletitle{Exploring larger evidence-base for contemporary Ayurveda}.
\newblock \bibinfo{journal}{\emph{International journal of Ayurveda research}} \bibinfo{volume}{1}, \bibinfo{number}{2} (\bibinfo{year}{2010}), \bibinfo{pages}{65}.
\newblock


\bibitem[\protect\citeauthoryear{S{\o}ndergaard and Hansen}{S{\o}ndergaard and Hansen}{2016}]%
        {sondergaard2016periodshare}
\bibfield{author}{\bibinfo{person}{Marie Louise~Juul S{\o}ndergaard} {and} \bibinfo{person}{Lone~Koefoed Hansen}.} \bibinfo{year}{2016}\natexlab{}.
\newblock \showarticletitle{PeriodShare: A bloody design fiction}. In \bibinfo{booktitle}{\emph{Proceedings of the 9th Nordic Conference on Human-Computer Interaction}}. \bibinfo{pages}{1--6}.
\newblock


\bibitem[\protect\citeauthoryear{S{\o}ndergaard, Kilic~Afsar, Ciolfi~Felice, Campo~Woytuk, and Balaam}{S{\o}ndergaard et~al\mbox{.}}{2020}]%
        {sondergaard2020designing}
\bibfield{author}{\bibinfo{person}{Marie Louise~Juul S{\o}ndergaard}, \bibinfo{person}{Ozgun Kilic~Afsar}, \bibinfo{person}{Marianela Ciolfi~Felice}, \bibinfo{person}{Nadia Campo~Woytuk}, {and} \bibinfo{person}{Madeline Balaam}.} \bibinfo{year}{2020}\natexlab{}.
\newblock \showarticletitle{Designing with Intimate Materials and Movements: Making" Menarche Bits"}. In \bibinfo{booktitle}{\emph{Proceedings of the 2020 ACM Designing Interactive Systems Conference}}. \bibinfo{pages}{587--600}.
\newblock


\bibitem[\protect\citeauthoryear{Spiel, Keyes, and Barlas}{Spiel et~al\mbox{.}}{2019}]%
        {spiel2019patching}
\bibfield{author}{\bibinfo{person}{Katta Spiel}, \bibinfo{person}{Os Keyes}, {and} \bibinfo{person}{P{\i}nar Barlas}.} \bibinfo{year}{2019}\natexlab{}.
\newblock \showarticletitle{Patching gender: Non-binary utopias in HCI}. In \bibinfo{booktitle}{\emph{Extended abstracts of the 2019 CHI conference on Human Factors in Computing Systems}}. \bibinfo{pages}{1--11}.
\newblock


\bibitem[\protect\citeauthoryear{Strauss and Corbin}{Strauss and Corbin}{1990}]%
        {75strauss1990open}
\bibfield{author}{\bibinfo{person}{Anselm Strauss} {and} \bibinfo{person}{Juliet Corbin}.} \bibinfo{year}{1990}\natexlab{}.
\newblock \showarticletitle{Open coding}.
\newblock \bibinfo{journal}{\emph{Basics of qualitative research: Grounded theory procedures and techniques}} \bibinfo{volume}{2}, \bibinfo{number}{1990} (\bibinfo{year}{1990}), \bibinfo{pages}{101--121}.
\newblock


\bibitem[\protect\citeauthoryear{Sultana}{Sultana}{2023}]%
        {sultana2023computing}
\bibfield{author}{\bibinfo{person}{Sharifa Sultana}.} \bibinfo{year}{2023}\natexlab{}.
\newblock \emph{\bibinfo{title}{Computing for Recognition: Design and Development of Just Technologies with Marginalized Communities}}.
\newblock \bibinfo{thesistype}{Ph.D. Dissertation}. \bibinfo{school}{Cornell University}.
\newblock


\bibitem[\protect\citeauthoryear{Sultana and Ahmed}{Sultana and Ahmed}{2019}]%
        {sultana2019witchcraft}
\bibfield{author}{\bibinfo{person}{Sharifa Sultana} {and} \bibinfo{person}{Syed~Ishtiaque Ahmed}.} \bibinfo{year}{2019}\natexlab{}.
\newblock \showarticletitle{Witchcraft and HCI: Morality, Modernity, and Postcolonial Computing in Rural Bangladesh}. In \bibinfo{booktitle}{\emph{Proceedings of the 2019 SIGCHI Conference on Human Factors in Computing Systems}}. ACM.
\newblock


\bibitem[\protect\citeauthoryear{Sultana, Ahmed, and Fussell}{Sultana et~al\mbox{.}}{2019}]%
        {sultana2019parar}
\bibfield{author}{\bibinfo{person}{Sharifa Sultana}, \bibinfo{person}{Syed~Ishtiaque Ahmed}, {and} \bibinfo{person}{Susan~R Fussell}.} \bibinfo{year}{2019}\natexlab{}.
\newblock \showarticletitle{"Parar-daktar Understands My Problems Better" Disentangling the Challenges to Designing Better Access to Healthcare in Rural Bangladesh}.
\newblock \bibinfo{journal}{\emph{Proceedings of the ACM on Human-Computer Interaction}} \bibinfo{volume}{3}, \bibinfo{number}{CSCW}, \bibinfo{pages}{1--27}.
\newblock


\bibitem[\protect\citeauthoryear{Sultana, Ahmed, and Rzeszotarski}{Sultana et~al\mbox{.}}{2021a}]%
        {sultana2020viz}
\bibfield{author}{\bibinfo{person}{Sharifa Sultana}, \bibinfo{person}{Syed~Ishtiaque Ahmed}, {and} \bibinfo{person}{Jeffrey~M Rzeszotarski}.} \bibinfo{year}{2021}\natexlab{a}.
\newblock \showarticletitle{Seeing in Context: Traditional Visual Communication Practices in Rural Bangladesh}.
\newblock \bibinfo{journal}{\emph{Proceedings of the ACM on Human-Computer Interaction}} \bibinfo{volume}{4}, \bibinfo{number}{CSCW3}, \bibinfo{pages}{1--31}.
\newblock


\bibitem[\protect\citeauthoryear{Sultana, Ahmed, and Rzeszotarski}{Sultana et~al\mbox{.}}{2023}]%
        {sultana2023communicating}
\bibfield{author}{\bibinfo{person}{Sharifa Sultana}, \bibinfo{person}{Syed~Ishtiaque Ahmed}, {and} \bibinfo{person}{Jeffrey~M Rzeszotarski}.} \bibinfo{year}{2023}\natexlab{}.
\newblock \showarticletitle{Communicating Consequences: Visual Narratives, Abstraction, and Polysemy in Rural Bangladesh}. In \bibinfo{booktitle}{\emph{Proceedings of the 2023 CHI Conference on Human Factors in Computing Systems}}. \bibinfo{pages}{1--19}.
\newblock


\bibitem[\protect\citeauthoryear{Sultana, Akter, Sultana, and Ahmed}{Sultana et~al\mbox{.}}{2022a}]%
        {sultana2022toleration}
\bibfield{author}{\bibinfo{person}{Sharifa Sultana}, \bibinfo{person}{Rokeya Akter}, \bibinfo{person}{Zinnat Sultana}, {and} \bibinfo{person}{Syed~Ishtiaque Ahmed}.} \bibinfo{year}{2022}\natexlab{a}.
\newblock \showarticletitle{Toleration Factors: The Expectations of Decorum, Civility, and Certainty on Rural Social Media}. In \bibinfo{booktitle}{\emph{Proceedings of the 2022 International Conference on Information and Communication Technologies and Development}}. \bibinfo{pages}{1--14}.
\newblock


\bibitem[\protect\citeauthoryear{Sultana, Deb, Bhattacharjee, Hasan, Alam, Chakraborty, Roy, Ahmed, Moitra, Amin, Islam, and Ahmed}{Sultana et~al\mbox{.}}{2021b}]%
        {sultana2021unmochon}
\bibfield{author}{\bibinfo{person}{Sharifa Sultana}, \bibinfo{person}{Mitrasree Deb}, \bibinfo{person}{Ananya Bhattacharjee}, \bibinfo{person}{Shaid Hasan}, \bibinfo{person}{SM~Raihanul Alam}, \bibinfo{person}{Trishna Chakraborty}, \bibinfo{person}{Prianka Roy}, \bibinfo{person}{Samira~Fairuz Ahmed}, \bibinfo{person}{Aparna Moitra}, \bibinfo{person}{M~Ashraful Amin}, \bibinfo{person}{A.K.M.~Najmul Islam}, {and} \bibinfo{person}{Syed~Ishtiaque Ahmed}.} \bibinfo{year}{2021}\natexlab{b}.
\newblock \showarticletitle{`Unmochon’: A Tool to Combat Online Sexual Harassment over Facebook Messenger}. In \bibinfo{booktitle}{\emph{Proceedings of the SIGCHI Conference on Human Factors in Computing Systems}}. \bibinfo{pages}{477--490}.
\newblock


\bibitem[\protect\citeauthoryear{Sultana and Fussell}{Sultana and Fussell}{2021}]%
        {sultana2021dissemination}
\bibfield{author}{\bibinfo{person}{Sharifa Sultana} {and} \bibinfo{person}{Susan~R Fussell}.} \bibinfo{year}{2021}\natexlab{}.
\newblock \showarticletitle{Dissemination, Situated Fact-checking, and Social Effects of Misinformation among Rural Bangladeshi Villagers During the COVID-19 Pandemic}.
\newblock \bibinfo{journal}{\emph{Proceedings of the ACM on Human-Computer Interaction}} \bibinfo{volume}{5}, \bibinfo{number}{CSCW2} (\bibinfo{year}{2021}), \bibinfo{pages}{1--34}.
\newblock


\bibitem[\protect\citeauthoryear{Sultana, Guimbreti{\`e}re, Sengers, and Dell}{Sultana et~al\mbox{.}}{2018}]%
        {sultana2018design}
\bibfield{author}{\bibinfo{person}{Sharifa Sultana}, \bibinfo{person}{Fran{\c{c}}ois Guimbreti{\`e}re}, \bibinfo{person}{Phoebe Sengers}, {and} \bibinfo{person}{Nicola Dell}.} \bibinfo{year}{2018}\natexlab{}.
\newblock \showarticletitle{Design Within a Patriarchal Society: Opportunities and Challenges in Designing for Rural Women in Bangladesh}. In \bibinfo{booktitle}{\emph{Proceedings of the 2018 CHI Conference on Human Factors in Computing Systems}}. ACM, \bibinfo{pages}{536}.
\newblock


\bibitem[\protect\citeauthoryear{Sultana, Islam, and Ahmed}{Sultana et~al\mbox{.}}{2020a}]%
        {sultana2020fighting}
\bibfield{author}{\bibinfo{person}{Sharifa Sultana}, \bibinfo{person}{AKM~Najmul Islam}, {and} \bibinfo{person}{Syed~Ishtiaque Ahmed}.} \bibinfo{year}{2020}\natexlab{a}.
\newblock \showarticletitle{Fighting coronavirus with faith: religious and parareligious responses to Covid-19 in Bangladesh}.
\newblock \bibinfo{journal}{\emph{Interactions}} \bibinfo{volume}{28}, \bibinfo{number}{1} (\bibinfo{year}{2020}), \bibinfo{pages}{6--7}.
\newblock


\bibitem[\protect\citeauthoryear{Sultana, Pritha, Tasnim, Das, Akter, Hasan, Alam, Kabir, and Ahmed}{Sultana et~al\mbox{.}}{2022b}]%
        {sultana2022shishushurokkha}
\bibfield{author}{\bibinfo{person}{Sharifa Sultana}, \bibinfo{person}{Sadia~Tasnuva Pritha}, \bibinfo{person}{Rahnuma Tasnim}, \bibinfo{person}{Anik Das}, \bibinfo{person}{Rokeya Akter}, \bibinfo{person}{Shaid Hasan}, \bibinfo{person}{SM~Raihanul Alam}, \bibinfo{person}{Muhammad~Ashad Kabir}, {and} \bibinfo{person}{Syed~Ishtiaque Ahmed}.} \bibinfo{year}{2022}\natexlab{b}.
\newblock \showarticletitle{‘ShishuShurokkha’: A Transformative Justice Approach for Combating Child Sexual Abuse in Bangladesh}. In \bibinfo{booktitle}{\emph{CHI Conference on Human Factors in Computing Systems}}. \bibinfo{pages}{1--23}.
\newblock


\bibitem[\protect\citeauthoryear{Sultana, Sultana, and Ahmed}{Sultana et~al\mbox{.}}{2020b}]%
        {sultana2020parareligious}
\bibfield{author}{\bibinfo{person}{Sharifa Sultana}, \bibinfo{person}{Zinnat Sultana}, {and} \bibinfo{person}{Syed~Ishtiaque Ahmed}.} \bibinfo{year}{2020}\natexlab{b}.
\newblock \showarticletitle{Parareligious-HCI: Designing for 'Alternative' Rationality in Rural Wellbeing in Bangladesh}. In \bibinfo{booktitle}{\emph{Extended Abstracts of the 2020 CHI Conference on Human Factors in Computing Systems}}. \bibinfo{pages}{1--13}.
\newblock


\bibitem[\protect\citeauthoryear{Svenningsen and Almeida}{Svenningsen and Almeida}{2020}]%
        {svenningsen2020designing}
\bibfield{author}{\bibinfo{person}{Ida~Kilias Svenningsen} {and} \bibinfo{person}{Teresa Almeida}.} \bibinfo{year}{2020}\natexlab{}.
\newblock \showarticletitle{Designing for the emotional pregnancy}. In \bibinfo{booktitle}{\emph{Companion Publication of the 2020 ACM Designing Interactive Systems Conference}}. \bibinfo{pages}{145--150}.
\newblock


\bibitem[\protect\citeauthoryear{Till, Farao, Coleman, Shandu, Khuzwayo, Muthelo, Mbombi, Bopane, Motlhatlhedi, Mabena, et~al\mbox{.}}{Till et~al\mbox{.}}{2022}]%
        {till2022community}
\bibfield{author}{\bibinfo{person}{Sarina Till}, \bibinfo{person}{Jaydon Farao}, \bibinfo{person}{Toshka~Lauren Coleman}, \bibinfo{person}{Londiwe~Deborah Shandu}, \bibinfo{person}{Nonkululeko Khuzwayo}, \bibinfo{person}{Livhuwani Muthelo}, \bibinfo{person}{Masenyani~Oupa Mbombi}, \bibinfo{person}{Mamare Bopane}, \bibinfo{person}{Molebogeng Motlhatlhedi}, \bibinfo{person}{Gugulethu Mabena}, {et~al\mbox{.}}} \bibinfo{year}{2022}\natexlab{}.
\newblock \showarticletitle{Community-based co-design across geographic locations and cultures: methodological lessons from co-design workshops in South Africa}. In \bibinfo{booktitle}{\emph{Proceedings of the Participatory Design Conference 2022-Volume 1}}. \bibinfo{pages}{120--132}.
\newblock


\bibitem[\protect\citeauthoryear{Truter}{Truter}{2007}]%
        {36truter2007african}
\bibfield{author}{\bibinfo{person}{Ilse Truter}.} \bibinfo{year}{2007}\natexlab{}.
\newblock \showarticletitle{African traditional healers: Cultural and religious beliefs intertwined in a holistic way}.
\newblock \bibinfo{journal}{\emph{South African Pharmaceutical Journal}} \bibinfo{volume}{74}, \bibinfo{number}{8} (\bibinfo{year}{2007}), \bibinfo{pages}{56--60}.
\newblock


\bibitem[\protect\citeauthoryear{Tuli, Chopra, Kumar, and Singh}{Tuli et~al\mbox{.}}{2018}]%
        {tuli2018learning}
\bibfield{author}{\bibinfo{person}{Anupriya Tuli}, \bibinfo{person}{Shaan Chopra}, \bibinfo{person}{Neha Kumar}, {and} \bibinfo{person}{Pushpendra Singh}.} \bibinfo{year}{2018}\natexlab{}.
\newblock \showarticletitle{Learning from and with Menstrupedia: Towards Menstrual Health Education in India}.
\newblock \bibinfo{journal}{\emph{Proceedings of the ACM on Human-Computer Interaction}} \bibinfo{volume}{2}, \bibinfo{number}{CSCW} (\bibinfo{year}{2018}), \bibinfo{pages}{174}.
\newblock


\bibitem[\protect\citeauthoryear{Tuli, Chopra, Singh, and Kumar}{Tuli et~al\mbox{.}}{2020}]%
        {tuli2020menstrual}
\bibfield{author}{\bibinfo{person}{Anupriya Tuli}, \bibinfo{person}{Shaan Chopra}, \bibinfo{person}{Pushpendra Singh}, {and} \bibinfo{person}{Neha Kumar}.} \bibinfo{year}{2020}\natexlab{}.
\newblock \showarticletitle{Menstrual (Im) mobilities and safe spaces}. In \bibinfo{booktitle}{\emph{Proceedings of the 2020 CHI Conference on Human Factors in Computing Systems}}. \bibinfo{pages}{1--15}.
\newblock


\bibitem[\protect\citeauthoryear{Tuli, Dalvi, Kumar, and Singh}{Tuli et~al\mbox{.}}{2019}]%
        {tuli2019sa}
\bibfield{author}{\bibinfo{person}{Anupriya Tuli}, \bibinfo{person}{Shruti Dalvi}, \bibinfo{person}{Neha Kumar}, {and} \bibinfo{person}{Pushpendra Singh}.} \bibinfo{year}{2019}\natexlab{}.
\newblock \showarticletitle{“It’sa girl thing” Examining Challenges and Opportunities around Menstrual Health Education in India}.
\newblock \bibinfo{journal}{\emph{ACM Transactions on Computer-Human Interaction (TOCHI)}} \bibinfo{volume}{26}, \bibinfo{number}{5} (\bibinfo{year}{2019}), \bibinfo{pages}{1--24}.
\newblock


\bibitem[\protect\citeauthoryear{Tuli, Singh, Narula, Kumar, and Singh}{Tuli et~al\mbox{.}}{2022}]%
        {tuli2022rethinking}
\bibfield{author}{\bibinfo{person}{Anupriya Tuli}, \bibinfo{person}{Surbhi Singh}, \bibinfo{person}{Rikita Narula}, \bibinfo{person}{Neha Kumar}, {and} \bibinfo{person}{Pushpendra Singh}.} \bibinfo{year}{2022}\natexlab{}.
\newblock \showarticletitle{Rethinking menstrual trackers towards period-positive ecologies}. In \bibinfo{booktitle}{\emph{Proceedings of the 2022 CHI Conference on Human Factors in Computing Systems}}. \bibinfo{pages}{1--20}.
\newblock


\bibitem[\protect\citeauthoryear{Tumpa, Islam, and Ankon}{Tumpa et~al\mbox{.}}{2017}]%
        {tumpa2017smart}
\bibfield{author}{\bibinfo{person}{Sanjida~Nasreen Tumpa}, \bibinfo{person}{Anika~Binte Islam}, {and} \bibinfo{person}{Md~Tasnim~Manzur Ankon}.} \bibinfo{year}{2017}\natexlab{}.
\newblock \showarticletitle{Smart care: An intelligent assistant for pregnant mothers}. In \bibinfo{booktitle}{\emph{2017 4th International Conference on Advances in Electrical Engineering (ICAEE)}}. IEEE, \bibinfo{pages}{754--759}.
\newblock


\bibitem[\protect\citeauthoryear{Tutia, Baljon, Vu, and Rosner}{Tutia et~al\mbox{.}}{2019}]%
        {tutia2019hci}
\bibfield{author}{\bibinfo{person}{Agatha Tutia}, \bibinfo{person}{Kelda Baljon}, \bibinfo{person}{Lan Vu}, {and} \bibinfo{person}{Daniela~K Rosner}.} \bibinfo{year}{2019}\natexlab{}.
\newblock \showarticletitle{HCI and menopause: Designing with and around the aging body}. In \bibinfo{booktitle}{\emph{Extended abstracts of the 2019 CHI conference on human factors in computing systems}}. \bibinfo{pages}{1--8}.
\newblock


\bibitem[\protect\citeauthoryear{Wailoo}{Wailoo}{2018}]%
        {wailoo2018historical}
\bibfield{author}{\bibinfo{person}{Keith Wailoo}.} \bibinfo{year}{2018}\natexlab{}.
\newblock \showarticletitle{Historical aspects of race and medicine: the case of J. Marion Sims}.
\newblock \bibinfo{journal}{\emph{Jama}} \bibinfo{volume}{320}, \bibinfo{number}{15} (\bibinfo{year}{2018}), \bibinfo{pages}{1529--1530}.
\newblock


\bibitem[\protect\citeauthoryear{Wall}{Wall}{2006}]%
        {wall2006medical}
\bibfield{author}{\bibinfo{person}{L~Lewis Wall}.} \bibinfo{year}{2006}\natexlab{}.
\newblock \showarticletitle{The medical ethics of Dr J Marion Sims: a fresh look at the historical record}.
\newblock \bibinfo{journal}{\emph{Journal of medical ethics}} \bibinfo{volume}{32}, \bibinfo{number}{6} (\bibinfo{year}{2006}), \bibinfo{pages}{346--350}.
\newblock


\bibitem[\protect\citeauthoryear{(WHO)}{(WHO)}{2021}]%
        {rephealth:online}
\bibfield{author}{\bibinfo{person}{World Health~Organization (WHO)}.} \bibinfo{year}{2021}\natexlab{}.
\newblock \bibinfo{howpublished}{\url{https://www.who.int/southeastasia/health-topics/reproductive-health}}.
\newblock


\bibitem[\protect\citeauthoryear{Wood, Wood, and Balaam}{Wood et~al\mbox{.}}{2017}]%
        {wood2017sex}
\bibfield{author}{\bibinfo{person}{Matthew Wood}, \bibinfo{person}{Gavin Wood}, {and} \bibinfo{person}{Madeline Balaam}.} \bibinfo{year}{2017}\natexlab{}.
\newblock \showarticletitle{Sex talk: Designing for sexual health with adolescents}. In \bibinfo{booktitle}{\emph{Proceedings of the 2017 Conference on Interaction Design and Children}}. \bibinfo{pages}{137--147}.
\newblock


\end{thebibliography}

\end{document}